\documentclass[11pt]{article}
\usepackage[pdftex]{color}
\usepackage{graphicx}
  \usepackage{amsthm,amsfonts}
  \usepackage{amsmath}
\newcommand{\bea}   {\begin{eqnarray}}
\newcommand{\eea}   {\end{eqnarray}}

\begin{document}
\renewcommand{\thefootnote}{\fnsymbol{footnote}}

\thispagestyle{empty}

\title{From worldline to quantum superconformal mechanics with/without oscillatorial terms: $D(2,1;\alpha) $ and $sl(2|1)$ models}

\author{I. E. Cunha\thanks{{E-mail: {\em ivanec@cbpf.br}}},\quad 
N. L. Holanda\thanks{{E-mail: {\em linneu@cbpf.br}}}
\quad and\quad F.
Toppan\thanks{{E-mail: {\em toppan@cbpf.br}}}
\\
\\
}
\maketitle

\centerline{
{\it CBPF, Rua Dr. Xavier Sigaud 150, Urca,}}{\centerline {\it\quad
cep 22290-180, Rio de Janeiro (RJ), Brazil.}
~\\
\maketitle
\begin{abstract}

In this paper we quantize superconformal $\sigma$-models defined by worldline supermultiplets. \par Two types of superconformal mechanics, with and without a DFF term, are considered. \par Without a DFF term (Calogero potential only) the supersymmetry is unbroken. \par The models with a DFF term correspond to deformed (if the Calogero potential is present) or undeformed oscillators. For these (un)deformed oscillators the classical invariant superconformal algebra acts as a spectrum-generating algebra of the quantum theory.\par
Besides the $osp(1|2)$ examples, we explicitly quantize the  superconformally-invariant worldine $\sigma$-models defined by the ${\cal N}=4$ $(1,4,3)$ supermultiplet (with $D(2,1;\alpha)$ invariance, for $\alpha\neq 0,-1$)   and by the ${\cal N}=2$ $(2,2,0)$ supermultiplet (with two-dimensional target and $sl(2|1)$ invariance). The parameter
$\alpha$ is the scaling dimension of the $(1,4,3)$ supermultiplet and, in the DFF case, has a direct interpretation as a vacuum energy. In the DFF case, for the $sl(2|1)$ models, the scaling dimension $\lambda$ is quantized (either $\lambda=\frac{1}{2}+{\mathbb Z}$ or $\lambda={\mathbb Z}$). The ordinary two-dimensional oscillator is recovered, after imposing a superselection restriction, from the  $\lambda=-\frac{1}{2}$ model. In particular a single bosonic vacuum is selected. The spectrum of the unrestricted two-dimensional theory is decomposed into an infinite set of lowest weight representations of $sl(2|1)$. Extra fermionic raising operators, not belonging to the original $sl(2|1)$ superalgebra, allow 
(for $\lambda=\frac{1}{2}+{\mathbb Z}$) to construct the whole spectrum from the two degenerate (one bosonic and one fermionic) vacua. 
~\\\end{abstract}
\vfill

\rightline{CBPF-NF-006/16
}

\newpage

\section{Introduction}

In this paper we quantize superconformal $\sigma$-models defined by worldline supermultiplets. We consider two types of superconformal mechanics, parabolic or trigonometric \cite{hoto}, namely in the absence or, respectively, in the presence of an oscillatorial DFF term \cite{dff}. \par
In the absence of a DFF term the systems under consideration possess only a Calogero potential \cite{cal}; they are supersymmetric and with a continuous spectrum. In the presence of a DFF term they correspond to deformed (if the Calogero potential is present) or undeformed oscillators with a discrete, bounded from below, spectrum. For these (un)deformed oscillators the classical invariant superconformal algebra acts as a spectrum-generating algebra of the quantum theory.\par
We illustrate at first our method with two $osp(1|2)$-invariant examples, the ordinary one-dimensional harmonic oscillator being recovered in the trigonometric case. Later we explicitly quantize the  superconformally-invariant worldine $\sigma$-models defined by\\
{\em i})  the ${\cal N}=4$ $(1,4,3)$ supermultiplet
with scaling dimension $\alpha\neq0,-1$ (these models are classically invariant under the exceptional $D(2,1;\alpha)$ Lie superalgebra) and \\
{\em ii}) the ${\cal N}=2$ $(2,2,0)$ supermultiplet of scaling dimension $\lambda$ (these models present a two-dimensional target and classical $sl(2|1)$-invariance). \par
For the $(1,4,3)$ supermultiplet, at the special $\alpha = -\frac{1}{2}$ value, the Calogero potential terms are vanishing. For this value the invariant superalgebra is $D(2,1;-\frac{1}{2})=D(2,1)\approx osp(4|2)$.\par
The results about the quantum parabolic $D(2,1;\alpha)$ models coincide with those obtained, with different methods, in \cite{fil2}. The new feature, in the present paper, is the construction of the quantum trigonometric models which, so far, have not been investigated. 
An interesting result, in the $(1,4,3)$ trigonometric case, consists in the direct and simple interpretation of $\alpha$ as a vacuum energy (if $\alpha$ is regarded as an external control parameter, it determines the Casimir energy of the system). \par
For the $sl(2|1)$ models the scaling dimension $\lambda$ is quantized (either $\lambda=\frac{1}{2}+{\mathbb Z}$ or $\lambda={\mathbb Z}$). In the trigonometric case the ordinary two-dimensional oscillator (without Calogero potential terms) is recovered from the special $\lambda=-\frac{1}{2}$ value after a superselection of the spectrum, defined by a projection operator, is imposed.  The restriction implies, in particular, that a single bosonic vacuum is obtained. The spectrum of the unrestricted theory turns out to be decomposed into an infinite set of lowest weight representations of $sl(2|1)$. By construction, the role of $sl(2|1)$ as a spectrum-generating algebra is expected. On the other hand, it is unexpected the further result that extra fermionic raising operators, not belonging to the $sl(2|1)$ superalgebra, allow to construct, for $\lambda=\frac{1}{2}+{\mathbb Z}$, the whole spectrum from the two degenerate (one bosonic and one fermionic) vacua  (in Appendix {\bf A} this action is visualized in diagrams).\par
Models of superconformal mechanics have been investigated in \cite{fr}--\cite{pap2} (see, e.g., the review  \cite{fil3} and references therein). For superconformal actions with oscillator potentials see \cite{{di},{bk},{hoto}}.  (Super)conformal mechanics is currently a very active area of research; among the motivations for 
this interest one can mention the $AdS_2/CFT_1$ correspondence \cite{{sen},{chaetal}}, or the possibility to apply it to test particles moving in the proximity of the horizon of certain black holes, see \cite{bmsv}.\par
${\cal N}=4$ superconformal models based on the exceptional (see \cite{dictionary}) Lie superalgebra $D(2,1;\alpha)$
were investigated  in \cite{ikl0}--\cite{khto}. The models considered in those works, mostly classical, are supersymmetric; for that reason they do not allow the presence of the oscillatorial DFF terms (in Appendix {\bf C} we comment about the ``soft" supersymmetry property of the oscillatorial models).  The recognition in \cite{pap} that conformal mechanics could allow new potentials, permitted the introduction in \cite{hoto} of the trigonometric (read, oscillatorial) classical $D(2,1;\alpha)$ models.

The scheme of the paper is the following.\par
Sections {\bf 2}, {\bf 3}, {\bf 4} are propaedeutic. In Section {\bf 2} we discuss the change of coordinates from linear to non-linear realizations of the superconformal algebras (the ``constant kinetic basis") which allows us to present the worldline superconformal $\sigma$-models in the Hamiltonian framework. A detailed description of the passage from classical Lagrangians to Hamiltonians is given in Section {\bf 3}. In Section {\bf 4} the quantization procedure and the construction of the Noether charges is explained for two examples, the  parabolic and trigonometric $osp(1|2)$-invariant $\sigma$-models. Section {\bf 5} contains the main results for the quantization of the parabolic (i.e. both superconformal and supersymmetric) quantum models with $D(2,1;\alpha)$-invariance, based on the ${\cal N}=4$ worldline supermultiplet $(1,4,3)$, and $sl(2,1)$-invariance, based on the ${\cal N}=2$ $ (2,2,0)$ worldline supermultiplet. In Section {\bf 6} the main results of their quantum trigonometric versions are derived. These systems contain DFF terms and are ``softly supersymmetric". They correspond to (un)deformed oscillators. The main results are the derivation of the vacuum energy in terms of the $\alpha$ scaling dimension for the $(1,4,3)$ supermultiplet
and the derivation of the spectrum-generating superalgebra  for the (un)deformed two-dimensional oscillator
with quantized scaling dimension $\lambda$.
In Appendix {\bf A} diagrams are presented illustrating the decomposition of the two-dimensional oscillators
in terms of the $sl(2|1)$ lowest weight representations, interconnected by the extra fermionic raising and lowering operators introduced in Section {\bf 6}.
For completeness in Appendix {\bf B} the classical version of the trigonometric ${\cal N}=2$ $(2,2,0)$ superconformal $\sigma$-model is presented. Finally, in Appendix {\bf C} we discuss the ``soft
supersymmetry" of the (un)deformed oscillators and the role, for these theories, of the spectrum generating superalgebras. In the Conclusions we present the open questions raised by our analysis.

\section{Worldline (super)conformal $\sigma$-models in constant kinetic basis}

A convenient approach, in constructing one-dimensional superconformal $\sigma$-models, consists in starting from a linear $D$-module representation of the superconformal algebra. Once such a representation is known, the Lagrangian defining the superconformally invariant action can be systematically constructed by applying fermionic generators  to a  prepotential function which depends only on the propagating bosons. The requirement of superconformal invariance, imposed as a constraint, determines the specific form of the prepotential.  This method (and its applications) has been discussed in \cite{hoto}. 

The kinetic term $\Phi({\vec x}) \frac{1}{2}\delta_{ij}({\dot x}_i{\dot x}_j+\ldots )$ of the derived Lagrangian is an ordinary constant kinetic term multiplied by a conformal factor $\Phi({\vec x})$ which is a function of the propagating bosons. In order to apply the standard methods of quantization we need to reabsorb the conformal factor. One way to do this consists in introducing a new set of fields. In the new basis of fields the kinetic term is expressed as a constant coefficient (hence the name ``constant kinetic basis" given in \cite{hoto}); the superalgebra, on the other hand, is realized non-linearly. 

In \cite{hoto} the procedure of changing the basis (from the ``linear" to the ``constant kinetic" basis) was sketched for certain $D$-module representations acting on supermultiplets consisting of a single propagating boson. We discuss it here in a more general framework.\par

Let us consider a $D$-module irrep of a $\mathcal{N}$-extended superconformal algebra (for our purposes ${\mathcal N}=1,2,4,8$)  acting on a $(k,\mathcal{N},\mathcal{N}-k)$ supermultiplet \cite{{pato}, {adi},{kuroto}, {ktref}} (namely, $k$ propagating bosons, ${\mathcal N}$ fermions and ${\mathcal N}-k$ bosonic auxiliary fields). In the linear basis the propagating bosons are labeled as $x_1$, ..., $x_k$, the fermions as $\psi_1$, ..., $\psi_{\mathcal{N}}$ and the auxiliary bosons as $b_1$, ..., $b_{\mathcal{N}-k}$. The kinetic term in the Lagrangian is given by
\begin{equation}
\label{linearkinetic}
\frac{1}{2}r^{-\frac{1+2\lambda}{\lambda}}(\dot{x}_m\dot{x}_m + i\omega\psi_{\beta}\dot{\psi}_\beta - \omega^2b_nb_n ).
\end{equation}
In the above equation the summation over the repeated indices is implied. The constant $\omega$ is dimensionless (and can be set equal to unity) in the parabolic case, while it is dimensional, see \cite{hoto},  in the hyperbolic/trigonometric case. The function $r$ is $r=(x_mx_m)^{\frac{1}{2}}$ and the parameter $\lambda$ is the scaling dimension of the supermultiplet. At $\lambda =-\frac{1}{2}$ the kinetic term is constant. For the remaining $\lambda\neq -\frac{1}{2}$ values a change to a constant kinetic basis is required in order to present a kinetic term with constant coefficients. Let us denote the propagating bosons in the constant kinetic basis as $y_1$, ..., $y_k$, the fermions as $\chi_1$, ..., $\chi_{\mathcal{N}}$ and the auxiliary bosons as $a_1$, ..., $a_{\mathcal{N}-k}$. The transformations passing from the ``linear" to the ``constant kinetic" basis are given by:\\
{\em i}) for the $(1,\mathcal{N},\mathcal{N}-1)$ supermultiplets we have
\bea
\label{1propbos}
y = -2\lambda x^{-\frac{1}{2\lambda}},\qquad 
\chi_{\beta} = x^{-\frac{1+2\lambda}{2\lambda}}\psi_{\beta}, \qquad a_n = x^{-\frac{1+2\lambda}{2\lambda}}b_n;
\eea
in terms of the new fields equation \eqref{linearkinetic} is expressed as
\bea
\label{constkin1propbos}
\frac{1}{2}(\dot{y}\dot{y} + i\omega\chi_{\beta}\dot{\chi}_\beta - \omega^2a_na_n );
\eea
{\em ii}) when $\mathcal{N}\geq 2$, for the $(2,\mathcal{N},\mathcal{N}-2)$ supermultiplets it is convenient to use a complex notation for the propagating bosons and set
\begin{align}
\label{2propbos}
y &= -2\lambda(x_1+ix_2)^{-\frac{1}{2\lambda}}, &y^* = -2\lambda(x_1-ix_2)^{-\frac{1}{2\lambda}}, \nonumber \\ 
\chi_{\beta} &= r^{-\frac{1+2\lambda}{2\lambda}}\psi_{\beta}, &a_n = r^{-\frac{1+2\lambda}{2\lambda}}b_n,\quad\quad\quad\quad
\end{align}
so that the kinetic term can be expressed as
\bea
\label{constkin2propbos}
\frac{1}{2}(\dot{y}\dot{y}^* + i\omega\chi_{\beta}\dot{\chi}_\beta - \omega^2a_na_n );
\eea
{\em iii}) when $\mathcal{N} = 4,8$ it is possible to construct a constant kinetic basis for any $(k,\mathcal{N},\mathcal{N}-k)$ supermultiplet at the specific $\lambda = 1/2$ value of the scaling dimension via the transformations
\bea
\label{lambda1/2basis}
y_m = \frac{x_m}{r^2}, \qquad \chi_\beta =  \frac{\psi_\beta}{r^2}, \qquad a_n =   \frac{b_n}{r^2},
\eea
leading to the kinetic term
\bea
\label{constkinlambda1/2}
\frac{1}{2}(\dot{y}_m\dot{y}_m +  i\omega\chi_{\beta}\dot{\chi}_\beta - \omega^2a_na_n ).
\eea
For $\mathcal{N} = 4$ and $k\neq 2$,
irreps of the exceptional superalgebras $D(2,1;\alpha)$ are recovered, see \cite{{kuto}, {khto},{hoto}}, from the $(k,4,4-k)$ supermultiplets according to the relation
\bea
\label{N=4-alpha-lambda-relation}
\alpha = (2-k)\lambda.
\eea 
At the special $\lambda = \frac{1}{2}$ value the associated superalgebra is $A(1,1)$ for the $(4,4,0)$ supermultiplet and $D(2,1)$ for the $(3,4,1)$ supermultiplet. 

For $\mathcal{N}=8$ and $k \neq 4$, irreps of superconformal algebras  are recovered for each supermultiplet $(k,8, 8-k)$ at the  critical values of the scaling dimension given by
\bea
\lambda_k = \frac{1}{k-4}.
\eea
The special value $\lambda=\frac{1}{2}$ yields an irrep of $A(3,1)$ acting on the supermultiplet $(6,8,2)$.
The reader is referred to \cite{{kuto},{khto}} for a detailed discussions on the criticality of the scaling dimension of the $\mathcal{N}=4,8$ superconformal algebras. 

\section{From Lagrangians to classical Hamiltonians: an application to the $osp(1|2)$-invariant $\sigma$-models}

The quantization of the 1D superconformal $\sigma$-models follows the canonical procedure formalized by Dirac and based on the classical Hamiltonian formalism. Since these $\sigma$-models have fermionic degrees of freedom, the passage from the Lagrangian to the classical Hamiltonian formalism requires the use of Dirac brackets (see, e.g., \cite{gity}). The need for Dirac brackets becomes clear after inspecting equations \eqref{constkin1propbos}, \eqref{constkin2propbos} and \eqref{constkinlambda1/2}; it is due to the fact that the linear dependence on the fermionic velocities $\dot{\chi}_\beta$ forces us to extend the phase space of the system and treat the fermionic canonical momenta as constraints in this extended phase space. In Dirac's language these constraints are both \emph{primary} (they hold even without using the equations of motion) and \emph{second class} (namely, a constraint that has a non-vanishing Poisson brackets with at least one of the constraints).   

This procedure, used throughout the paper, will be illustrated in detail for the simplest possibility  given by the $osp(1|2)$-invariant  $\sigma$-models (their two variants, \emph{parabolic} and  \emph{hyperbolic/trigonometric}, see \cite{hoto}). In the parabolic case the Hamiltonian is identified with a bosonic root of the superconformal algebra, while in the hyperbolic/trigonometric case it is associated with a  Cartan element. The parabolic $D$-module reps describe systems which are supersymmetric, while the  hyperbolic/trigonometric reps furnish only a {\em soft} version of supersymmetry, see the discussion in the Introduction. The hyperbolic and trigonometric models are interrelated via a Wick rotation of the dimensional parameter $\omega$. The trigonometric case is here emphasized with respect to the hyperbolic one because it yields a bounded from below  Hamiltonian.\par
In the rest of this Section we discuss in detail the Hamiltonian formulation of both parabolic and trigonometric $osp(1|2)$-invariant  $\sigma$-models.  The method, notations and conventions here presented are later applied to models with larger superconformal symmetry. \\

\subsection{The $osp(1|2)$-invariant parabolic $\sigma$-model}

In the constant kinetic basis the generators of the $osp(1|2)$ parabolic $D$-module rep read as
\bea
\label{osp1|2pargen}
&&H = \left(\begin{array}{cc}
\partial_t & 0\\
0 & \partial_t
\end{array}\right),\quad
D = \left(\begin{array}{cc}
t\partial_t - \frac{1}{2} & 0\\
0 & t\partial_t
\end{array}\right),\quad
K = \left(\begin{array}{cc}
t^2\partial_t -  t & 0\\
0 & t^2\partial_t
\end{array}\right),\nonumber\\
&&Q = \left(\begin{array}{cc}
0 & 1\\
i\partial_t & 0
\end{array}\right),\quad
\bar{Q} =\left( \begin{array}{cc}
0 & t\\
it\partial_t - i & 0
\end{array}\right).
\eea
The above generators act on the column vector supermultiplet $(y,\chi)^T$ possessing the scaling dimension $\lambda=-\frac{1}{2}$. \par The bosonic generators $H$, $D$, $K$ span the $sl(2)$ Lie subalgebra, while the fermionic generators $Q$, $\bar{Q}$ span the odd sector of $osp(1|2)$.  \par
The associated $osp(1|2)$-invariant action is simply
\bea 
\label{osp1|2parlagr}
&{\mathcal S}=\int dt \mathcal{L} = \int dt\frac{1}{2}(\dot{y}^2 + i\chi\dot{\chi}).&
\eea
Unlike the ${\cal N}\geq 2$ superconformal algebras discussed in the following, for $osp(1|2)$ the same action is recovered by starting from a generic $D$-module rep with scaling dimension $\lambda\neq -\frac{1}{2}$ and applying the (\ref{1propbos}) change of basis.\par
For a theory possessing bosons and fermions a conserved Noether charge is expressed, for a symmetry generator $O$, as
\bea
\label{noethercharges}
C_O =  (\delta_O\phi_I)\dfrac{\partial \mathcal{L}}{\partial \dot{\phi}_I}-J_O,
\eea
where $J_O$ stems from the variation  $\delta_O \mathcal{L} = \frac{dJ_O}{dt}$; the sum over the repeated index $I$ labeling the fields is understood. The given ordering of the right hand side of  \eqref{noethercharges} is essential in dealing with Grassmann variables and derivatives.  \par
For the case at hand the classical Noether charges are
\bea\label{N=1parclasschar}
&C_H = \frac{\dot{y}^2}{2},\quad C_D = \frac{t\dot{y}^2}{2} - \frac{y\dot{y}}{2},\quad C_K = \frac{t^2\dot{y}^2}{2} - ty\dot{y} + \frac{y^2}{2} , \quad C_Q = \dot{y}\chi,\quad C_{\bar{Q}} = t\dot{y}\chi + y\chi.&
\eea
The Euler-Lagrange equations
\bea
\label{eulerlagrange}
\frac{\partial\mathcal{L}}{\partial\phi} = \frac{d}{dt}(\frac{\partial\mathcal{L}}{\partial\dot{\phi}})
\eea
lead to the equations of motion
\bea
\label{N=1pareqofmotion}
\ddot{y} = 0, \qquad \dot{\chi} =0.
\eea
The Grassmann variable in the classical $osp(1|2)$ model is a constant and plays essentially no physical role besides ensuring the $osp(1|2)$ invariance. \par
To introduce the Hamiltonian formalism we have to compute the conjugate momenta given by 
\bea
\label{N=1parconjmomenta}
&p = \frac{\partial \mathcal{L}}{\partial\dot{y}} = \dot{y}, \quad
\pi =  \frac{\partial \mathcal{L}}{\partial\dot{\chi}} = -\frac{i\chi}{2}.&
\eea
In the Hamiltonian framework the classical charges \eqref{N=1parclasschar} are rewritten as
\bea\label{N=1hamparclasschar}
C_H = \frac{p^2}{2},\quad C_D = \frac{tp^2}{2} - \frac{yp}{2},\quad C_K = \frac{t^2p^2}{2} - typ + \frac{y^2}{2}, \quad
C_Q = p\chi,\quad C_{\bar{Q}} = tp\chi + y\chi.
\eea
The last step requires defining the Dirac brackets. The second equation in \eqref{N=1parconjmomenta} makes clear why Dirac brackets need to be introduced.  The conjugate momentum $\pi$ to the Grassmann variable $\chi$ is not an invertible function of the velocity $\dot{\chi}$. The second equation in \eqref{N=1parconjmomenta} should therefore be viewed as a second class constraint on the phase space, 
\bea
\label{N=1parconstraint}
u = \pi + \frac{i\chi}{2}.
\eea
The super-Poisson bracket involving even or odd $f$, $g$ functions is given by 
\bea
\label{superpoissonbrackets}
\{f,g\}_P = \sum_I (-1)^{deg(f)\cdot deg(g)}\frac{\partial f}{\partial \phi_I}\frac{\partial g}{\partial \pi_I} - \frac{\partial f}{\partial \pi_I}\frac{\partial g}{\partial \phi_I},
\eea
where the degree function $deg$ is $0$ if evaluated on bosons and $1$ on fermions. \par
Denoting with $u_i$ the set of all second class contraints, the Dirac bracket reads as 
\bea
\label{diracbrackets}
\{f,g\}_D = \{f,g\}_P - \sum_{k,l}\{f,u_k\}_PU^{-1}_{kl}\{u_l,g\}_P,
\eea
where $U_{kl} = \{u_k,u_l\}_P$ is a matrix constructed from the super-Poisson brackets of all second class constraints. \par
$u$ entering \eqref{N=1parconstraint} is a second class constraint, since it satisfies
\begin{equation*}
\{u,u\}_P = -i.
\end{equation*}
A straightforward computation gives the non-vanishing Dirac brackets
\bea
\label{N=1pardiracbrackets}
\{y,p\}_D = 1,\qquad \{\chi,\chi\}_D = -i.
\eea
We can derive, with the use of the Dirac brackets, the equations of motion in the Hamiltonian formalism and compute (recovering $osp(1|2)$) the superalgebra satisfied by the \eqref{N=1hamparclasschar} conserved charges.  \par In terms of Dirac brackets the Hamilton's equations are
\bea
\label{parhamilton}
\dot{\phi} =  \frac{\partial\phi}{\partial t} +  \{\phi,C_H\}_D.
\eea
For the case at hand we get
\bea
\label{N=1parhamiltoneq}
\dot{p} = 0,\qquad \dot{\chi} = 0,
\eea
which, together with the $p={\dot y}$ position, allow to recover \eqref{N=1pareqofmotion}.  \\ 

\subsection{The $osp(1|2)$-invariant trigonometric $\sigma$-model}

In the trigonometric case the passage from the Lagrangian to the Hamiltonian formalism follows the same steps as before. We therefore skip unnecessary comments.\par In the constant kinetic basis the generators of the $osp(1|2)$ trigonometric $D$-module rep are 
\bea
\label{osp1|2triggen}
&&H = e^{i\omega t}\left(\begin{array}{cc}
\frac{1}{\omega}\partial_t -\frac{i}{2}  & 0\\
0 & \frac{1}{\omega}\partial_t
\end{array}\right),\quad
D = \left(\begin{array}{cc}
\frac{1}{\omega}\partial_t& 0\\
0 & \frac{1}{\omega}\partial_t
\end{array}\right),\quad
K =  e^{-i\omega t}\left(\begin{array}{cc}
\frac{1}{\omega}\partial_t +\frac{i}{2} & 0\\
0 & \frac{1}{\omega}\partial_t 
\end{array}\right),\nonumber\\
&&Q = e^{\frac{i\omega t}{2}}\left(\begin{array}{cc}
0 & 1\\
\frac{i}{\omega}\partial_t + \frac{1}{2}  & 0
\end{array}\right),\qquad
\bar{Q} = e^{-\frac{i\omega t}{2}}\left(\begin{array}{cc}
0 & 1\\
\frac{i}{\omega}\partial_t - \frac{1}{2}  & 0
\end{array}\right).
\eea
The $osp(1|2)$-invariant action is
\bea 
\label{osp1|2hyplagr}
{\mathcal S}=\int dt \mathcal{L} = \int dt\frac{1}{2}(\dot{y}^2 + i\omega\chi\dot{\chi}- \frac{\omega^2}{8}y^2).
\eea
The derived conserved Noether charges are 
\bea &&
C_H = e^{i\omega t}(\frac{1}{2\omega}\dot{y}^2 - \frac{i}{2}y\dot{y}-\frac{\omega}{8}y^2),\quad C_D = \frac{1}{2\omega}\dot{y}^2 + \frac{\omega}{8}y^2, \quad C_K =  e^{-i\omega t}(\frac{1}{2\omega}\dot{y}^2 + \frac{i}{2}y\dot{y}-\frac{\omega}{8}y^2),\nonumber\\
&&
\label{N=1trigclasschar}
C_Q = e^{\frac{i\omega}{2}t}(\dot{y}\chi - \frac{i\omega}{2}y\chi),\quad  C_{\bar{Q}} =  e^{-\frac{i\omega}{2}t}(\dot{y}\chi + \frac{i\omega}{2}y\chi).
\eea
The Euler-Lagrange equations of motion are 
\bea
\label{N=1trigeqofmotion}
\ddot{y} = -\frac{\omega^2y}{4}, \qquad \dot{\chi} =0.
\eea
The conjugate momenta are given by
\bea
\label{N=1trigconjmomenta}
&p = \frac{\partial \mathcal{L}}{\partial\dot{y}} = \dot{y}, \quad
\pi =  \frac{\partial \mathcal{L}}{\partial\dot{\chi}} = -\frac{i\omega\chi}{2}.&
\eea
In the Hamiltonian formulation, the \eqref{N=1trigclasschar} conserved charges are
\bea
&&C_H = e^{i\omega t}(\frac{1}{2\omega}p^2 - \frac{i}{2}yp-\frac{\omega}{8}y^2),\quad C_D = \frac{1}{2\omega}p^2 + \frac{\omega}{8}y^2, \quad C_K =  e^{-i\omega t}(\frac{1}{2\omega}p^2 + \frac{i}{2}yp-\frac{\omega}{8}y^2),\nonumber\\
\label{N=1hamtrigclasschar}
&&C_Q = e^{\frac{i\omega}{2}t}(p\chi - \frac{i\omega}{2}y\chi),\quad  C_{\bar{Q}} =  e^{-\frac{i\omega}{2}t}(p\chi + \frac{i\omega}{2}y\chi).
\eea
The second equation in \eqref{N=1trigconjmomenta} gives the constraint in phase space
\begin{equation}
\label{N=1trigconstraint}
u = \pi + \frac{i\omega\chi}{2},
\end{equation}
which allows to compute the Dirac brackets as before. The non-vanishing Dirac brackets are
\bea
\label{N=1trigdiracbrackets}
\{y,p\}_D = 1,\qquad \{\chi,\chi\}_D = -\frac{i}{\omega}.
\eea
The Hamilton's equations of motion are now written as 
\bea
\label{trighamilton}
\dot{\phi} = \omega\{\phi,C_D\}_D + \frac{\partial\phi}{\partial t}.
\eea
One should note that, while in the parabolic $\sigma$-model the charge $C_H$ is the physical Hamiltonian and the symmetry operator $H$ is the generator of the time translations, in the trigonometric $\sigma$-model  the physical hamiltonian is given by $\omega C_D$, the Cartan generator $\omega D$ being the generator of the time translations. 
One can readily check that equation \eqref{trighamilton} leads to 
\bea
\label{N=1parhamiltoneqtr}
\dot{p} = -\frac{\omega^2y}{4},\qquad \dot{\chi} = 0,
\eea 
which reproduce \eqref{N=1trigeqofmotion} by taking into account that $p={\dot y}$. 

\section{The quantization. Quantum versus classical Noether charges and the $osp(1|2)$ models}

The canonical quantization of the models presented in Section {\bf 3} is realized by substituting the Dirac Brackets by the appropriate (based on the superalgebra structure) (anti)commutators, that we will denote with the ``$[.,.\}$" symbol:
\bea
\label{canonicalquantization}
\{A,B\}_D &\rightarrow &\frac{1}{i\hbar}[A,B\}.
\eea
By applying \eqref{canonicalquantization} to \eqref{N=1pardiracbrackets} and \eqref{N=1trigdiracbrackets} we get, respectively,  the parabolic and trigonometric $osp(1|2)$-invariant quantum superconformal models.\par
 We point out  that, since the observables must be Hermitian operators, the parabolic and trigonometric quantum models correspond to different real forms (read, conjugations) of the invariant superalgebra. We illustrate in detail this  feature, which is also valid for ${\cal N}\geq 2$ invariant theories.

\subsection{The parabolic $osp(1|2)$-invariant quantum $\sigma$-model}

The non-vanishing (anti)commutators recovered from \eqref{N=1pardiracbrackets} are
\bea
\label{N=1parsuperbrackets}
[\hat{y},\hat{p}] = i\hbar, &&\{\hat{\chi},\hat{\chi}\} = \hbar.
\eea
In the position-space representation the above operators are given by
\bea
\label{N=1parquantumoperators}
&\hat{y} = y,\qquad \hat{p}=-i\hbar\partial_y,\qquad \hat{\chi} = \sqrt{\frac{\hbar}{2}}.&
\eea
The last equation is particularly important because it tells us that the fermionic field $\chi$, classically represented by a Grassmann variable,  becomes a Clifford variable $\hat{\chi}$ in the quantum version.
The choice in (\ref{N=1parquantumoperators}) of representing $\hat{\chi}$ as a real number is not unique. An alternative choice, which respects the ${\mathbb Z}_2$-graded structure of the super-vector space acted upon by the operators
${\hat y}, {\hat p}, {\hat\chi}$, consists in picking ${\hat \chi}$ as the $2\times 2$ matrix $\sqrt{\frac{\hbar}{2}}\left(\begin{array}{cc} 0&1\\1&0\end{array}\right)$. In this ${\mathbb Z}_2$-graded representation, the operators ${\hat y}, {\hat p}, {\hat\chi}$ are
\bea\label{z2preserving}
&\hat{y} = \left(\begin{array}{cc} y&0\\ 0&y\end{array}\right),\quad \hat{p}=
\left(\begin{array}{cc} -i\hbar\partial_y&0\\ 0&-i\hbar\partial_y\end{array}\right),\quad \hat{\chi} =
 \sqrt{\frac{\hbar}{2}}\left(\begin{array}{cc} 0&1\\ 1&0\end{array}\right),\quad 
N_f = \left(\begin{array}{cc} 1&0\\ 0&-1\end{array}\right),
&
\eea
while $N_f$ is the Fermion Parity operator.\par
The possibility, offered by the ${\mathbb Z}_2$-graded structure, of doubling the vector space, will be used in the following in constructing  ${\cal N}=2, 4$ quantum models. \par
It is worth pointing out that superalgebras admit super-representations acting on ${\mathbb Z}_2$-graded vector spaces. In some cases superalgebra (anti)commutation relations are also realized on ordinary (not ${\mathbb Z}_2$-graded) vector spaces. This feature can be seen when realizing the 
$\chi^2={\mathbb I}$ equation either through $\chi =1$ or the $\chi =\left(\begin{array}{cc} 0&1\\1&0\end{array}\right)$ ${\mathbb Z}_2$-graded solution (they induce a $ Cl(1,0)$ Clifford algebra 
which is respectively identified either with $Cl(1,0)\approx {\mathbb R}$ or with the split-complex numbers $Cl(1,0)\approx {\widetilde{\mathbb C}}$). 
Upon a convenient normalization, equation (\ref{N=1parquantumoperators}) corresponds to the first choice, while equation (\ref{z2preserving}) corresponds to the second choice.\par
It is worth pointing out that the different quantum models derived from equations  (\ref{N=1parquantumoperators}) and (\ref{z2preserving}) (only the latter one being supersimmetric) are both consistent. The (\ref{N=1parquantumoperators}) model can be derived from the (\ref{z2preserving}) model after imposing a superselection rule induced by a projector (a similar projector inducing a superselection rule is introduced in Appendix {\bf A}). For simplicity we discuss in this Section the parabolic (and its trigonometric counterpart, see equation (\ref{quantumtrigosp12charges})) model corresponding to the first choice. The ${\mathbb Z}_2$-graded choice is used in Sections {\bf 5} and {\bf 6} to derive ${\cal N}=4$ and ${\cal N}=2$ quantum models.
\par
The (\ref{quantumtrigosp12charges}) model coincides with the ordinary quantum oscillator (its connection with the $osp(1|2)$ superalgebra is elucidated in Appendix {\bf C}).
\par
The parabolic quantum $osp(1|2)$ superalgebra obtained by the \eqref{canonicalquantization} quantization of the classical counterpart, leads to
\bea
&&[\hat{H},\hat{D}] = i\hbar \hat{H}, \quad~ [\hat{H},\hat{K}] = 2i\hbar\hat{D}, \quad [\hat{K},\hat{D}] = -i\hbar \hat{K} \nonumber\\
&& [\hat{H},\hat{\bar{Q}}] = i\hbar \hat{Q}, \quad ~~ [\hat{K},\hat{Q}] = -i\hbar \hat{\bar{Q}}, \quad [\hat{Q},\hat{D}] = \frac{i\hbar}{2}\hat{Q}, \quad [\hat{\bar{Q}},\hat{D}] = -\frac{i\hbar}{2}\hat{\bar{Q}}, \nonumber\\
\label{quantumparosp1|2}
&&\{\hat{Q},\hat{Q}\} = 2\hbar  \hat{H}, \quad \{\hat{Q},\hat{\bar{Q}}\} = 2\hbar D, \quad \{\hat{\bar{Q}},\hat{\bar{Q}}\} = 2\hbar K.
\eea
The remaining (anti)commutators are vanishing. \par
The above  superalgebra is realized by the quantum charges
\bea
&&\hat{H} = \frac{1}{2}\hat{p}^2, \quad \hat{D} = \frac{t}{2}\hat{p}^2 - \frac{1}{4}(\hat{y}\hat{p} + \hat{p}\hat{y}), \quad \hat{K} = \frac{t^2}{2}\hat{p}^2 - \frac{t}{2}(\hat{y}\hat{p}+\hat{p}\hat{y}) + \frac{1}{2}\hat{y}^2, \nonumber\\
\label{quantumparosp1|2charges}
&&\hat{Q} = \hat{\chi}\hat{p}, \quad ~~ \hat{\bar{Q}} = t\hat{\chi}\hat{p} - \hat{y}\hat{\chi}.
\eea

They are, up to symmetrization, identical to the classical charges. This is a unique feature of the $\mathcal{N} =1$ $osp(1|2)$-invariant models. From  $\mathcal{N}\geq 2$ the models explicitly depend on the scaling dimension $\lambda$. As a result, the quantum versions of these theories require corrections which are traced backed to the mapping of the classical Grassmann variables into quantum Clifford generators. \par

The Hamiltonian $\hat{H}$  in \eqref{quantumparosp1|2charges} corresponds to the one-dimensional free particle. The operators $\hat{H}$, $\hat{D}$ $\hat{K}$ close the $sl(2)$ bosonic symmetry algebra of the system. $\hat{H}$ and $\hat{Q}$ gives the ${\mathcal N}=1$ algebra of the Supersymmetric Quantum Mechanics. In terms of the (\ref{N=1parquantumoperators}) realization ($\hat{\chi}$ is a real number) the parabolic $osp(1|2)$-invariant model admits no fermionic degrees of freedom. This is no longer the case (fermions are present) if the model is expressed via the (\ref{z2preserving}) realization.  \par   
In the parabolic model all charges entering \eqref{quantumparosp1|2charges} are observables. The superalgebra \eqref{quantumparosp1|2} can be re-expressed in terms of the canonical $osp(1|2)$ Cartan-Weyl basis $H, F^\pm, E^\pm$ (such that all the structure constants are real),
see \cite{dictionary}, through the identifications
\bea
\label{parrealCWbasis}
&\hat{H} =- E^-, \quad \hat{D} = i H, \quad \hat{K} = - E^+, \quad \hat{Q} = 2F^-, \quad \hat{\bar{Q}} = 2i F^+.&
\eea
The computation of the $osp(1|2)$ structure constants in the new basis is immediate.\par
The superalgebra conjugation corresponding to (\ref{quantumparosp1|2charges}) reads, in the Cartan-Weyl basis, as
\bea
\label{conjpar}
&(E^{\pm})^{\dagger} = E^{\pm}, \quad H^{\dagger} = -H, \quad (F^{\pm})^{\dagger} = \mp (F^{\pm}).&
\eea

Concerning the dimensional analysis of the model we can set, without loss of generality, $[\partial_t] = 1$.
If we set the Planck constant $\hbar$ and the action ${\cal S}$ to be dimensionless, we therefore get
$[\hat{y}] = -\frac{ 1}{2}$, $[\hat{p}] = \frac{1}{2}$, $[\hat{\chi}] =[{\cal S}]=0.$

\subsection{The trigonometric $osp(1|2)$-invariant quantum $\sigma$-model}

The quantization of the trigonometric model follows the same lines of the parabolic one. Without loss of generality we can set $\omega =1$, reproducing the non-vanishing (anti)commutators (\ref{N=1parsuperbrackets}) and the
(\ref{N=1parquantumoperators}) and (\ref{z2preserving}) position-space representations for the operators $\hat{y},\hat{p}, {\hat \chi}$.\\
The quantum trigonometric generators, identical to the classical ones up to symmetrization, are 
\bea
\label{quantumtrigosp12charges}
&&\hat{H} = e^{it}(\frac{1}{2}\hat{p}^2 -\frac{i}{4}(\hat{y}\hat{p}+\hat{p}\hat{y}) - \frac{1}{8}\hat{y}^2), \quad \hat{K} = e^{-i t}(\frac{1}{2}\hat{p}^2 + \frac{i}{4}(\hat{y}\hat{p}+\hat{p}\hat{y}) - \frac{}{8}\hat{y}^2), \nonumber\\
&&\hat{D} = \frac{1}{2}\hat{p}^2 + \frac{1}{8}\hat{y}^2, \quad ~~\hat{Q} =e^{\frac{i t}{2}}(\hat{\chi}\hat{p} - \frac{i}{2}\hat{\chi}\hat{y}), \quad ~~\hat{\bar{Q}} =e^{-\frac{i t}{2}}(\hat{\chi}\hat{p} + \frac{i}{2}\hat{\chi}\hat{y}).
\eea 
In the (\ref{quantumtrigosp12charges}) realization, the $osp(1|2)$ non-vanishing brackets read as
\bea
&&~[\hat{H},\hat{D}] = \hbar \hat{H}, ~\quad [\hat{H},\hat{K}] = 2\hbar\hat{D}, ~\quad [\hat{K},\hat{D}] = -\hbar \hat{K} ,\nonumber\\
&&~[\hat{H},\hat{\bar{Q}}] = \hbar \hat{Q}, ~~\quad [\hat{K},\hat{Q}] = -\hbar \hat{\bar{Q}}, ~\quad [\hat{Q},\hat{D}] = \frac{\hbar}{2}\hat{Q}, \quad [\hat{\bar{Q}},\hat{D}] = -\frac{\hbar}{2}\hat{\bar{Q}}, \nonumber\\
\label{quantumtrigosp1|2}
&&\{\hat{Q},\hat{Q}\} = 2\hbar  \hat{H}, \quad \{\hat{Q},\hat{\bar{Q}}\} = 2\hbar D, \quad \{\hat{\bar{Q}},\hat{\bar{Q}}\} = 2\hbar K.
\eea
The $osp(1|2)$ Cartan-Weyl basis is recovered, from the (\ref{quantumtrigosp12charges}) trigonometric charges, via the identifications
\bea
\label{trigrealCWbasis}
&\hat{H} =  E^-, \quad \hat{D} = H, \qquad \hat{K} = - E^+, \quad \hat{Q} = 2iF^-, \quad \hat{\bar{Q}} = -2i F^+.&
\eea
We obtain a different conjugation with respect to the parabolic case, given by
\bea
\label{conjtrig}
(E^{\pm})^{\dagger} = -E^{\mp}, \qquad H^{\dagger} = H, \qquad (F^{\pm})^{\dagger} = F^{\mp}.
\eea
In the trigonometric case the Hamiltonian is given by the $osp(1|2)$ Cartan generator $ \omega\hat{D}$.\par
By taking into account the presence of the dimensional parameter $\omega$ that we set, for convenience, equal to $1$ in the formulas above, the dimensional analysis of the trigonometric model gives us the dimensions
$[t]=-1$, $ [\hat{y}] = -\frac{ 1}{2}$, $[\hat{p}] = \frac{1}{2}$, $[\hat{\chi}] =-\frac{1}{2}$, $[\omega]=1$, $[{\cal S}]=0$.\par

\section{Superconformal Quantum Mechanics with Calogero potentials: $1D$ $D(2,1;\alpha)$ and $2D$ $sl(2|1)$ models}

In this Section we quantize the worldline superconformal $\sigma$-models recovered from the $\mathcal{N}=4$ $(1,4,3)$
(i.e., one-dimensional target) and $\mathcal{N}=2$ $(2,2,0)$ (i.e., two-dimensional target) parabolic supermultiplets. Unlike the $\mathcal{N} =1$ parabolic model analyzed in Section {\bf 4}, non-trivial potential terms and non-trivial quantum corrections to the classical Hamiltonians, appear. 

The  $\mathcal{N}=4$ $(1,4,3)$ parabolic model possesses a $D(2,1;\alpha)$ invariance, where $\alpha\neq 0,-1$ is identified with the scaling dimension of the supermultiplet. The Hamiltonian describes a particle moving on a line under an inverse square potential and includes spin-like degrees of freedom.   

The $\mathcal{N}=2$ $(2,2,0)$ parabolic model possesses an $sl(2|1)$ invariance. Its Hamiltonian describes a particle moving on a plane under an inverse square potential and with a spin-orbit coupling.

\subsection{The $\mathcal{N}=4$ $(1,4,3)$ parabolic model with $D(2,1;\alpha)$ invariance}

A discussion of the classical $\mathcal{N}=4$ $(1,4,3)$ superconformal worldline model can be found, e.g., in \cite{hoto}. We present here the quantization of this model repeating the same steps discussed in Section {\bf 4} for the $osp(1|2)$-invariant model. In this subsection we recover, within a different framework, the models discussed in \cite{fil2}.\par
The non-vanishing (anti)commutators obtained from quantizing the Dirac brackets are  
\bea
\label{N=4parsuperbrackets}
&[\hat{y},\hat{p}] = i,\qquad \{\hat{\chi}_{\alpha},\hat{\chi}_{\beta}\} = \delta_{\alpha\beta},&
\eea 
with $\alpha,\beta = 0,...,3$. The above equations define the superalgebra $\mathfrak{h}_1\oplus C_4$, with  the one-dimensional Heisenberg algebra $\mathfrak{h}_1$ in its even sector and the four $C\ell(4,0)$ Clifford algebra gamma-matrices in its odd sector. These gamma-matrices can be expressed as $4\times 4$ complex matrices. We choose, to respect the $\mathbb{Z}_2$-graded structure of the superalgebra,  block-antidiagonal gamma matrices, while representing the Heisenberg generators as block-diagonal operators. \par
The position-space representation of \eqref{N=4parsuperbrackets} is
\bea
&&~~\hat{y} = y\mathbb{I}_4, \quad \quad\hat{p} = -i\partial_y\mathbb{I}_4,\quad\nonumber\\~
\label{N=4parquantumoperators}
&&\hat{\chi}_0 = \frac{1}{\sqrt{2}}\sigma_2\otimes\mathbb{I}_2, \quad \hat{\chi}_1 = -\frac{1}{\sqrt{2}}\sigma_1\otimes\sigma_1, \quad \hat{\chi}_2 = -\frac{1}{\sqrt{2}}\sigma_1\otimes\sigma_2, \quad \hat{\chi}_3 = -\frac{1}{\sqrt{2}}\sigma_1\otimes\sigma_3,
\eea
where $\mathbb{I}_n$ is the $n$ x $n$ identity matrix and the $\sigma_i$'s ($i=1,2,3$) are the Pauli matrices. \par 
The quantum charges are given by 
\bea
\label{N=4parcharges}
\hat{H} &=& (\frac{\hat{p}^2}{2} + \frac{(1+2\alpha)^2}{8\hat{y}^2})\mathbb{I}_4 + \frac{1+2\alpha}{4\hat{y}^2}\mathcal{F}_4, \nonumber\\
\hat{D} &=& (\frac{t\hat{p}^2}{2} -\frac{1}{4}(\hat{y}\hat{p} + \hat{p}\hat{y}) + \frac{t(1+2\alpha)^2}{8\hat{y}^2})\mathbb{I}_4 + \frac{t(1+2\alpha)}{4\hat{y}^2}\mathcal{F}_4, \nonumber\\
\hat{K} &=& (\frac{t^2\hat{p}^2}{2} -\frac{t}{2}(\hat{y}\hat{p} + \hat{p}\hat{y}) + \frac{\hat{y}^2}{2} + \frac{t^2(1+2\alpha)^2}{8\hat{y}^2})\mathbb{I}_4 + \frac{t^2(1+2\alpha)}{4\hat{y}^2}\mathcal{F}_4,\nonumber\\
\hat{Q}_0 &=& \hat{\chi}_0\hat{p} + \frac{i(1+2\alpha)}{6}\epsilon_{ijk}\frac{\hat{\chi}_i\hat{\chi}_j\hat{\chi}_k}{\hat{y}}, \nonumber\\
\hat{Q}_i &=& \hat{\chi}_i\hat{p} - \frac{i(1+2\alpha)}{2}\epsilon_{ijk}\frac{\hat{\chi}_0\hat{\chi}_j\hat{\chi}_k}{\hat{y}}, \nonumber\\
\hat{\bar{Q}}_0 &=& t\hat{\chi}_0\hat{p} - \chi_0\hat{y} + \frac{it(1+2\alpha)}{6}\epsilon_{ijk}\frac{\hat{\chi}_i\hat{\chi}_j\hat{\chi}_k}{\hat{y}}, \nonumber\\
\hat{\bar{Q}}_i &=& t\hat{\chi}_i\hat{p} - \chi_i\hat{y} - \frac{it(1+2\alpha)}{2}\epsilon_{ijk}\frac{\hat{\chi}_0\hat{\chi}_j\hat{\chi}_k}{\hat{y}}, \nonumber\\
\hat{J}_i &=& -i(\frac{1}{2}\epsilon_{ijk}\hat{\chi}_j\hat{\chi}_k + \hat{\chi}_0\hat{\chi}_i), \nonumber\\
\hat{L}_i &=& -i(\frac{1}{2}\epsilon_{ijk}\hat{\chi}_j\hat{\chi}_k - \hat{\chi}_0\hat{\chi}_i).
\eea
In the above formulas we used the Fermi Parity operator $\mathcal{F}_4$, defined by 
$\mathcal{F}_{2n} = {\scriptsize{\left(\begin{array}{cc}
\mathbb{I}_n & 0\\
0 & -\mathbb{I}_n
\end{array}\right)}}
$.
\\
One should note that the quantum operators $\hat{H}$, $ \hat{D}$, $\hat{K}$ contain an Ehrenfest quantum correction term, proportional to $\frac{\hbar^2(1+2\alpha)^2}{\hat{y}^2}\mathbb{I}_4$, which is not present in the classical charges. Its appearance can be traced to the change from classical Grassmann to quantum Clifford variables.\par
  At a given value $\alpha\neq 0, -1$, the above operators close the exceptional superalgebra $D(2,1;\alpha)$. The R-symmetry generators $\hat{J}_i$ and $\hat{L}_i$, $i=1,2,3$, close two independent 
($[\hat{J}_i,\hat{L}_j] = 0$) $su(2)$ subalgebras.\par
In the Cartan-Weyl basis the non-vanishing $D(2,1;\alpha)$ brackets are given by 
\bea
&&~~~[H,E^{\pm}] = \pm E^{\pm}, \quad [E^+,E^-] = 2H, \quad [H,F^{\pm}_\beta] = \pm \frac{1}{2}F^{\pm}_\beta, \quad[E^{\pm},F^{\mp}_\beta] = -F^{\pm}_\beta, \nonumber \\
&&\{F^{\pm}_0,F^{\mp}_j\} = -\frac{i}{4}(\lambda J_j + (1+\lambda)L_j), \quad \{F_j^+,F_k^-\} = \epsilon_{jkl}(-\frac{i\lambda}{4}J_l + \frac{i(\lambda+1)}{4}L_l)+\delta_{jk}\frac{H}{2},\nonumber \\
&&\{F_0^+,F_0^-\}= \frac{H}{2},\quad\{F^{\pm}_\beta,F^{\pm}_\gamma\} = \pm \frac{1}{2}\delta_{\beta\gamma} E^{\pm},\quad [J_j,F^{\pm}_0] = iF^{\pm}_j,\quad [J_j,F^{\pm}_k] = i(-\delta_{jk}F^{\pm}_0 + \epsilon_{jkl}F^{\pm}_l),\nonumber\\
&&~~~[L_j,F^{\pm}_0] = -iF^{\pm}_j,~\quad \quad[L_j,F^{\pm}_k] = i(\delta_{jk}F^{\pm}_0 + \epsilon_{jkl}F^{\pm}_l),\nonumber\\
\label{D(2,1;alpha)}
&&~~~~ [J_j,J_k] = 2i\epsilon_{jkl}J_l, ~\quad\quad [L_j,L_k] = 2i\epsilon_{jkl}L_l. 
\eea
The above superalgebra is realized by the \eqref{N=4parcharges} quantum operators via the identifications 
\bea
&\hat{H} =- E^-, \quad \hat{D} = i H, \quad \hat{K} =- E^+, \quad \hat{Q}_\beta = 2F^-_\beta, \quad \hat{\bar{Q}}_\beta = 2i F^+_\beta, \quad
\label{N=4parbasis}
\hat{J}_j = J_j, \quad \hat{L}_j = L_j.&
\eea
 The Hamiltonian operator $\hat{H}$, explicitly written in $4\times 4$ supermatrix form,  is given by 
\begin{equation}
\label{N=4calogeromodel}
\hat{H} = 
\left(
\begin{array}{c|c}
  (\frac{\hat{p}^2}{2} + \frac{4\alpha^2 + 8\alpha + 3}{8\hat{y}^2})\mathbb{I}_2  & 0 \\ \hline
  0 & (\frac{\hat{p}^2}{2} + \frac{4\alpha^2 -1}{8\hat{y}^2})\mathbb{I}_2
\end{array}
\right).
\end{equation}
It is the Hamiltonian of the $\mathcal{N} =4$ super-Calogero model with $D(2,1;\alpha)$ invariance.\par
It contains a (purely bosonic) Calogero Hamiltonian in both its upper and lower diagonal blocks. We recall that the Calogero Hamiltonian ${\mathcal H}_{C}$ is given by
\begin{equation}
\label{calogeromodel}
{\mathcal{H}}_{C} = \frac{1}{2}\hat{p}^2 + \frac{g^2}{\hat{y}^2}.
\end{equation} 
The self-adjointness of the Calogero Hamiltonian ${\mathcal{H}}_{C}$ depends on the value of the coupling parameter $g$. We refer to the \cite{{olpe2},{and}}  papers for a thorough discussion of this subtle point.\par
For our purposes it is important to note here the relation between the coupling constant $g$ and the scaling dimension parameter $\alpha$. From \cite{olpe2} we know that ${\mathcal{H}}_{C}$ is self-adjoint, provided that the inequality $g^2 > -\frac{1}{8}$ is satisfied. Under this condition the boundary value problem 
\begin{equation*}
{\mathcal{H}}_{C}\phi_k = E_k\phi_k, \qquad \phi_k(0) = 0,
\end{equation*} 
gives a continuous positive spectrum, $0\leq E_k < \infty$, the eigenfunctions and eigenvalues being
\begin{equation*}
\phi_k(y) = 2^{\mu-\frac{1}{2}}\Gamma(\mu+\frac{1}{2})(ky)^{-(\mu -\frac{1}{2})}J_{\mu-\frac{1}{2}}(ky)y^\mu, \qquad E_k = \frac{1}{2}k^2,
\end{equation*}
for
\begin{equation}
\label{g-mu-relation-par}
g^2 = \frac{1}{2}\mu(\mu-1).
\end{equation}
Let us set
\bea
\label{g-alpha-relation-par}
g_b^2 =  \frac{4\alpha^2 + 8\alpha + 3}{8}, && g_f^2 = \frac{4\alpha^2 -1}{8},
\eea
for the Calogero parameters entering, respectively,  the upper and lower diagonal blocks of the
\eqref{N=4calogeromodel} Hamiltonian. It is quite rewarding that imposing, simultaneously, the $g_b^2, g_f^2 >- \frac{1}{8}$ condition, we end up with the $\alpha \neq 0,-1$ inequality for the scaling dimension.
The class of exceptional $D(2,1;\alpha)$ superalgebras guarantee the existence of a well-defined Hamiltonian
with a continuous positive spectrum bounded from below.\par
At the special $\alpha = -\frac{1}{2}$ value the Calogero potential terms (in both upper and lower blocks) vanish. Therefore, this special point corresponds to a free theory. At this given value, see \cite{dictionary}, we have $D(2,1;-\frac{1}{2})=D(2,1)$, so that the invariant superalgebra coincides with the classical $D(2,1)\approx osp(4|2)$ superalgebra. \par
We can express, from \eqref{g-mu-relation-par}, $g_b,g_f$ in terms of their respective $\mu_b,\mu_f$ parameters. From \eqref{g-alpha-relation-par} $\mu_b$, $\mu_f$ can be given in terms of $\alpha$. The result is the linear relations
\bea
\label{mu-alpha-relation-par}
&\mu_b = \frac{1}{2}\pm(\alpha+1), \quad \mu_f = \frac{1}{2}\pm \alpha.&
\eea
In quantum mechanics the continuity conditions are also imposed on the probability currents. Since the zero-energy wave function is (up to a normalizing factor) $\phi_0(y) = y^\mu$, these conditions imply that both $\mu_b, \mu_f$ must satisfy $\mu_b,\mu_f>\frac{1}{2}$ to ensure continuity at the origin. The \eqref{mu-alpha-relation-par} equations show that any $\alpha \neq 0,-1$ is suitable to fulfill these constraints. \par
As a final comment we point out that the energy levels of both bosonic (upper)  and fermionic (lower) blocks are doubly degenerated.  This degeneracy is removed by taking into account the hermitian operators $\hat{J}_3$, $\hat{L}_3$ which commute with $\hat{H}$. Indeed,
\bea
\label{N=4calogero-j3-k3}
&\hat{J}_3 = 
\left(
\begin{array}{c|c}
  \sigma_3  & 0 \\ \hline
  0 & 0
\end{array}
\right), \qquad \hat{L}_3 = 
\left(
\begin{array}{c|c}
  0  & 0 \\ \hline
  0 & \sigma_3
\end{array}
\right),&
\eea
are both diagonal and specify spin-like quantum numbers in the bosonic and fermionic sectors, respectively. We can say that the bosonic states have $\frac{1}{2}$ $\hat{J}$-spin  and $0$ $\hat{L}$-spin, while the fermionic states have $0$ $\hat{J}$-spin and $\frac{1}{2}$ $\hat{L}$-spin .

\subsection{The $\mathcal{N}=2$ $(2,2,0)$ parabolic model with $sl(2|1)$ invariance}

The classical $sl(2|1)$-invariant action based on the parabolic D-module rep of the $\left(2,2,0\right)$ supermultiplet is presented in Appendix {\bf B}. Its quantization is performed with the techniques previously outlined (introduction of the ``constant kinetic basis", Dirac brackets, etc.). For this model it is convenient to express the two propagating bosons in terms of a complex field $y$. \par
We obtain the non-vanishing (anti)commutators
\bea
&[y^{*},p_{y^{*}}]=[y,p_y]=i\hbar,\quad\{ \chi,\chi^{\dagger}\} =\frac{\hbar}{C},&
\eea
\\
where $p_{y}=-i\hbar\partial_{y}$, $p_{y^{*}}=-i\hbar\partial_{y^{*}}$ and the fermions can be expressed as $\chi=\sqrt{\frac{\hbar}{C}}\left(\begin{array}{cc}
0 & 1\\
0 & 0
\end{array}\right)$ and $\chi^{\dagger}=\sqrt{\frac{\hbar}{C}}\left(\begin{array}{cc}
0 & 0\\
1 & 0
\end{array}\right)$. \\
Let us fix, for simplicity, $\hbar=1$ and $C=\frac{1}{2}$. Then the quantum charges can be written as

\bea
\hat{H} &=& (2p_{y}p_{y^{*}}+\frac{(2\lambda +1)^{2}}{8yy^{*}})\mathbb{I}_2+i\frac{2\lambda +1}{4}(\chi\chi^{\dagger}-\chi^{\dagger}\chi)(\frac{p_{y^{*}}}{y}-\frac{p_y}{y^{*}}), \nonumber\\
\hat{D} &=& t\hat{H}-\frac{1}{2}(y^{*}p_{y^{*}}+yp_{y}-i)\mathbb{I}_2, \nonumber\\
\hat{K} &=& t^2\hat{H}-t(y^{*}p_{y^{*}}+yp_{y}-i)\mathbb{I}_2+\frac{1}{2}yy^{*}\mathbb{I}_2,\nonumber\\
{\hat{Q}^{(1)}}_{-} &=& -\frac{i}{2}\left(\left(\frac{y}{y^{*}}\right)^{\frac{1+2\lambda}{2}}p_{y} + p_{y}\left(\frac{y}{y^{*}}\right)^{\frac{1+2\lambda}{2}}\right)\chi - \frac{i}{2}\left(\left(\frac{y^{*}}{y}\right)^{\frac{1+2\lambda}{2}}p_{y^{*}} + p_{y^{*}}\left(\frac{y^{*}}{y}\right)^{\frac{1+2\lambda}{2}}\right)\chi^{\dagger}, \nonumber\\
{\hat{Q}^{(2)}}_{-}  &=& -\frac{1}{2}\left(\left(\frac{y}{y^{*}}\right)^{\frac{1+2\lambda}{2}}p_{y} + p_{y}\left(\frac{y}{y^{*}}\right)^{\frac{1+2\lambda}{2}}\right)\chi + \frac{1}{2}\left(\left(\frac{y^{*}}{y}\right)^{\frac{1+2\lambda}{2}}p_{y^{*}} + p_{y^{*}}\left(\frac{y^{*}}{y}\right)^{\frac{1+2\lambda}{2}}\right)\chi^{\dagger}, \nonumber\\
{\hat{Q}^{(1)}}_{+} &=& t{\hat{Q}^{(1)}}_{-} - \frac{1}{\sqrt{2}}\sqrt{yy^{*}}\left(\left(\frac{y}{y^{*}}\right)^{\lambda}\chi  + \left(\frac{y^{*}}{y}\right)^{\lambda}\chi^{\dagger}\right), \nonumber\\
{\hat{Q}^{(2)}}_{+} &=& t{\hat{Q}^{(2)}}_{-} - \frac{i}{\sqrt{2}}\sqrt{yy^{*}}\left(\left(\frac{y}{y^{*}}\right)^{\lambda}\chi  - \left(\frac{y^{*}}{y}\right)^{\lambda}\chi^{\dagger}\right), \nonumber\\
\hat{J} &=& \frac{i}{2}(\frac{p_{y^{*}}}{y}-\frac{p_y}{y^{*}})-\frac{1-2\lambda}{8}(\chi\chi^{\dagger}-\chi^{\dagger}\chi).
\eea
\\
Here $\hat{H}$ is the quantum hamiltonian.

Using $p_{y}=-i\hbar\partial_{y}$, $p_{y^{*}}=-i\hbar\partial_{y^{*}}$, the quantum operators $\hat{Q}_{-}^{(1)}$, $\hat{Q}_{-}^{(2)}$ turn out to be

\bea
\hat{Q}_{-}^{(1)}=i{\footnotesize{\left(\begin{array}{cc}
0 & -A\\
A^{\dagger} & 0
\end{array}\right)}}, &&\hat{Q}_{-}^{(2)}={\footnotesize{\left(\begin{array}{cc}
0 & A\\
A^{\dagger} & 0
\end{array}\right)}},
\eea
where
\bea
&A^{\dagger}=-\frac{i}{\sqrt{2}}e^{-i2\lambda\theta}\left(\partial_{r}+\frac{i}{r}\partial_{\theta}+\frac{2\lambda+1}{2r}\right),\quad A=-\frac{i}{\sqrt{2}}e^{i2\lambda\theta}\left(\partial_{r}-\frac{i}{r}\partial_{\theta}+\frac{2\lambda+1}{2r}\right)\label{eq:AAdagger}
\eea
\\
are expressed in polar coordinates ($y=re^{i\theta}$, $y^{*}=re^{-i\theta}$).\par
In the same way, the quantum hamiltonian $\hat{H}$ can be expressed as
\bea
&\hat{H}=[-\frac{1}{2}\left(\partial_{r}^{2}+\frac{1}{r}\partial_{r}+\frac{1}{r^{2}}\partial_{\theta}^{2}\right)+i\frac{\left(2\lambda+1\right)}{2r^{2}}\sigma_{3}\partial_{\theta}+\frac{\left(2\lambda+1\right)^{2}}{8r^{2}}]\mathbb{I}_2,\label{eq:hamiltonian220p}&
\eea
\\
with $\sigma_{3}$ being the diagonal Pauli matrix. $\frac{\left(2\lambda+1\right)^{2}}{8r^{2}}$ is the Ehrenfest term resulting from quantization.

The non-vanishing (anti)commutators, closing the $sl(2|1)$ superalgebra are ($m,n=0,\pm 1$):
\bea
&&[\hat{L}_{n},\hat{L}_{m}]=i\left(m-n\right)\hat{L}_{m+n},\quad [\hat{L}_{0},\hat{Q}_{\pm}^{I}]=\pm\frac{i}{2}\hat{Q}_{\pm}^{I},\quad [\hat{L}_{\pm1},\hat{Q}_{\mp}^{I}]=\mp i\hat{Q}_{\pm}^{I},\nonumber\\
&&[\hat{J},\hat{Q}_{\pm}^{I}]=\frac{i}{2}\epsilon_{IJ}\hat{Q}_{\pm}^{J},\quad\{ \hat{Q}_{\pm}^{I},\hat{Q}_{\pm}^{J}\} =2\delta_{IJ}\hat{L}_{\pm 1},\quad \{ \hat{Q}_{\pm}^{I},\hat{Q}_{\mp}^{J}\} =2\delta_{IJ}\hat{L}_{0} \pm 2\epsilon_{IJ}\hat{J},\nonumber\\&&
\eea
\\
where $\hat{L}_{-1}=\hat{H}$, $\hat{L}_{0}=\hat{D}$, $\hat{L}_{1}=\hat{K}$, $I,J=1,2$ and $\epsilon_{12}=-\epsilon_{21}=1$.

The eigenvalue equation $\hat{H}\psi_{E_{m\pm}}=E_{m\pm}\psi_{E_{m\pm}}$,
for $E_{m\pm}>0$, produces a continuum spectrum with eigenfunctions
\begin{eqnarray}
\psi_{Em+}(r,\theta) & = & J_{|\frac{2\lambda+1}{2}-m|}(\alpha r)e^{im\theta}{\footnotesize{\left(\begin{array}{c}
1\\
0
\end{array}\right)}},\nonumber \\
\psi_{Em-}(r,\theta) & = & J_{|\frac{2\lambda+1}{2}+m|}(\alpha r)e^{im\theta}{\footnotesize{\left(\begin{array}{c}
0\\
1
\end{array}\right)}},
\end{eqnarray}
\\
where $J_{|\frac{2\lambda+1}{2}-m|}(\alpha r)$
and $J_{|\frac{2\lambda+1}{2}+m|}(\alpha r)$
are Bessel functions and $\alpha=\sqrt{2E}$.\par
To conclude the analysis of this model we present it as a Supersymmetric Quantum Mechanics. Let us introduce
\bea
\hat{Q}=\frac{\hat{Q}_{-}^{2}+i\hat{Q}_{-}^{1}}{2}={\footnotesize{\left(\begin{array}{cc}
0 & A\\
0 & 0
\end{array}\right)}}, &&\hat{Q}^{\dagger}=\frac{\hat{Q}_{-}^{2}-i\hat{Q}_{-}^{1}}{2}={\footnotesize{\left(\begin{array}{cc}
0 & 0\\
A^{\dagger} & 0
\end{array}\right)}}.
\eea
We get $\{ \hat{Q},\hat{Q}^{\dagger}\} =2\hat{H}$ and $\hat{Q}^{2}=(\hat{Q}^{\dagger})^{2}=0$. \par 
From the expressions (\ref{eq:AAdagger}), it follows that $\hat{Q}\psi_{E_{m-}}=\psi_{E_{\left(m+2\lambda\right)+}}$
and $\hat{Q}^{\dagger}\psi_{E_{m+}}=\psi_{E_{\left(m-2\lambda\right)-}}$.
Since $m+2\lambda$ and $m-2\lambda$ need to be integer numbers, $\hat{Q}\psi_{E_{m-}}$
and $\hat{Q}^{\dagger}\psi_{E_{m+}}$ belong to the Hilbert 
space only if $2\lambda$ is an integer number. A supersymmetric pair is therefore only encountered for the quantized values of the scaling dimension, either $\lambda\in \frac{1}{2}+{\mathbb Z}$ or $\lambda\in{\mathbb Z}$.

\section{Superconformal Quantum Mechanics with DFF oscillator potential terms: $1D$ $D(2,1;\alpha)$ and $2D$ $sl(2|1)$ models }

In this Section we quantize the worldline trigonometric $\sigma$-models obtained from the
$\mathcal{N}=4$ $(1,4,3)$ and $\mathcal{N}=2$ $(2,2,0)$ supermultiplets (see Appendix {\bf B}). They contain (besides a Calogero potential) an oscillatorial (DFF) term which furnishes a discrete, bounded from below, spectrum. The associated $D(2,1;\alpha)$ and, respectively, $sl(2|1)$superconformal algebras act as spectrum-generating algebras for these models.\par

The $D(2,1;\alpha)$ $(1,4,3)$ trigonometric $\sigma$-models  shed some new light on the 
results of Calogero \cite{cal} and de Alfaro, Fubini and Furlan \cite{dff}. Indeed, their Casimir energy linearly depends (in two regions) on the scaling dimension parameter $\alpha$  
(in contrast with the complicated dependence expressed in terms of the Calogero coupling constant, see \cite{olpe2}).\par

For what concerns the $sl(2|1)$ $(2,2,0)$ trigonometric $\sigma$-models interesting features are also obtained. The scaling dimension $\lambda$ needs to be quantized (either $\lambda =\frac{1}{2}+{\mathbb Z}$ or $\lambda\in {\mathbb Z}$). At the special $\lambda = -\frac{1}{2}$ value the ordinary two-dimensional oscillator (since the Calogero potential vanishes at this special point) can be recovered after performing a restriction induced by a superselection rule. The restriction selects, in particular, a single bosonic vacuum. The Hilbert space of the unrestricted two-dimensional models is decomposed into an infinite direct sum of $sl(2|1)$ lowest weight representations.
An unexpected feature is the existence of fermionic raising operators (not entering the $sl(2|1)$ superalgebra) which allow, together with the $sl(2|1)$ raising operators, for $\lambda =\frac{1}{2}+{\mathbb Z}$ to recover the whole Hilbert space of the theory from the two degenerate (one bosonic and one fermionic) vacua of the theory. The existence of these extra fermionic operators is traced to the presence of a discrete symmetry.

\subsection{The quantum $D(2,1;\alpha)$ trigonometric model from $\mathcal{N}=4$ $(1,4,3)$}

The quantization of this model follows the same steps as the quantization of the $osp(1|2)$-invariant trigonometric model described in Section {\bf 4}. We end up, just like its $\mathcal{N} = 4$ $(1,4,3)$ parabolic counterpart of Section {\bf 5}, with (anti)commutators defining the the $\mathfrak{h}_1 \oplus C_4$ superalgebra \eqref{N=4parsuperbrackets}. We set, for convenience and without loss of generality, the dimensional parameter $\omega=1$ (its presence in the equations can be restored by means of dimensional analysis). \par
The quantum operators are ($\mathcal{F}_4$ is the Fermion Parity operator introduced in (\ref{N=4parcharges}))
\bea
\label{N=4trigcharges}
\hat{H} &=& e^{it}(\frac{\hat{p}^2}{2} -\frac{i}{4}(\hat{y}\hat{p}+\hat{p}\hat{y}) - \frac{\hat{y}^2}{8}+ \frac{(1+2\alpha)^2}{8\hat{y}^2})\mathbb{I}_4 + e^{it}\frac{1+2\alpha}{4\hat{y}^2}\mathcal{F}_4, \nonumber\\
\hat{D} &=& (\frac{\hat{p}^2}{2} + \frac{\hat{y}^2}{8} + \frac{(1+2\alpha)^2}{8\hat{y}^2})\mathbb{I}_4 + \frac{(1+2\alpha)}{4\hat{y}^2}\mathcal{F}_4, \nonumber\\
\hat{K} &=& e^{-it}(\frac{\hat{p}^2}{2} +\frac{i}{4}(\hat{y}\hat{p}+\hat{p}\hat{y}) - \frac{\hat{y}^2}{8}+ \frac{(1+2\alpha)^2}{8\hat{y}^2})\mathbb{I}_4 + e^{-it}\frac{1+2\alpha}{4\hat{y}^2}\mathcal{F}_4,\nonumber\\
\hat{Q}_0 &=& e^{\frac{it}{2}}(\hat{\chi}_0\hat{p} - \frac{i}{2}\hat{\chi}_0{\hat{y}} + \frac{i(1+2\alpha)}{6}\epsilon_{ijk}\frac{\hat{\chi}_i\hat{\chi}_j\hat{\chi}_k}{\hat{y}}), \nonumber\\
\hat{Q}_i &=& e^{\frac{it}{2}}(\hat{\chi}_i\hat{p} - \frac{i}{2}\hat{\chi}_i{\hat{y}} - \frac{i(1+2\alpha)}{2}\epsilon_{ijk}\frac{\hat{\chi}_0\hat{\chi}_j\hat{\chi}_k}{\hat{y}}), \nonumber\\
\hat{\bar{Q}}_0 &=& e^{-\frac{it}{2}}(\hat{\chi}_0\hat{p} + \frac{i}{2}\hat{\chi}_0{\hat{y}} + \frac{i(1+2\alpha)}{6}\epsilon_{ijk}\frac{\hat{\chi}_i\hat{\chi}_j\hat{\chi}_k}{\hat{y}}), \nonumber\\
\hat{\bar{Q}}_i &=& e^{-\frac{it}{2}}(\hat{\chi}_i\hat{p} + \frac{i}{2}\hat{\chi}_i{\hat{y}} - \frac{i(1+2\alpha)}{2}\epsilon_{ijk}\frac{\hat{\chi}_0\hat{\chi}_j\hat{\chi}_k}{\hat{y}}), \nonumber\\
\hat{J}_i &=& -i(\frac{1}{2}\epsilon_{ijk}\hat{\chi}_j\hat{\chi}_k + \hat{\chi}_0\hat{\chi}_i), \nonumber\\
\hat{L}_i &=& -i(\frac{1}{2}\epsilon_{ijk}\hat{\chi}_j\hat{\chi}_k - \hat{\chi}_0\hat{\chi}_i).
\eea
The above operators realize the $D(2,1;\alpha)$ superalgebra \eqref{D(2,1;alpha)} with the identifications
\bea
&\hat{H} =  E^-, \quad \hat{D} = H, \quad \hat{K} =- E^+, \quad \hat{Q}_\beta =2iF^-_\beta, \quad \hat{\bar{Q}}_\beta =-2i F^+_\beta, \quad
\label{N=4parbasis}
\hat{J}_j = J_j, \quad \hat{L}_j = L_j.&
\eea
The quantum Hamiltonian $\hat{\mathcal{H}} \equiv\hat{D}$ is, explicitly,
\bea
\label{N=4dffmodel}
\hat{D}& =& 
\left(
\begin{array}{c|c}
  (\frac{\hat{p}^2}{2} + \frac{4\alpha^2 + 8\alpha + 3}{8\hat{y}^2} + \frac{\hat{y}^2}{8})\mathbb{I}_2  & 0 \\ \hline
  0 & (\frac{\hat{p}^2}{2} + \frac{4\alpha^2 -1}{8\hat{y}^2} + \frac{\hat{y}^2}{8})\mathbb{I}_2
\end{array}
\right).
\eea
Both upper (bosonic) and lower (fermionic) diagonal blocks of $\hat{D}$ contain a Calogero Hamiltonian with the DFF oscillatorial potential, 
\begin{equation}
\label{dffmodel}
\hat{\mathcal{H}}_{DFF} = \frac{1}{2}\hat{p}^2 + \frac{g^2}{\hat{y}^2} + \frac{\hat{y}^2}{8}.
\end{equation} 
%H = \begin{bmatrix}
%\partial_t & 0\\
%0 & \partial_t
%\end{bmatrix}
%\end{equation}
A detailed analysis of this Hamiltonian can be found in \cite{{cal},{olpe2}}. Just like the parabolic case, the inequality $g^2 > -\frac{1}{8}$ guarantees the existence of physically acceptable solutions. The boundary value problem
\bea
&\hat{\mathcal{H}}_{DFF}\phi_n = E_n\phi_n, \quad \phi_n(0) = 0,\qquad n=0,1,2,\ldots,&
\eea
implies the discrete spectrum
\begin{equation}
\label{143trigenergies}
E_n =\frac{1}{2}(n + \nu + 1), 
\end{equation}
with eigenfunctions given (up to normalization) by 
\begin{equation}
\label{143trigeigenfuctions}
\phi_n(y) = y^{\nu+\frac{1}{2}}\exp(-\frac{y^2}{4})L^\nu_n(\frac{1}{2}y^2).
\end{equation}
In the right hand side $L^\nu_n$ stands for the modified Laguerre polynomials. The parameter $\nu$ entering the Casimir energy $\frac{1}{2}(\nu +1)$ is
\bea
\label{g-nu-relation-trig}
\nu = \frac{1}{2}(1+8g^2)^{\frac{1}{2}}.
\eea
Comparing equations \eqref{N=4dffmodel} and \eqref{dffmodel} we see that $g_b$, $g_f$ are again given by equations \eqref{g-alpha-relation-par}, so that $\alpha\neq 0,-1$ to ensure that both $g_b^2$ and $g_f^2$ are greater than $-\frac{1}{8}$. \par
Since the Hamiltonian is a Cartan generator of the (\ref{N=4trigcharges}) superalgebra, the whole spectrum can be recovered from a lowest weight representation of $D(2,1;\alpha)$, where the $Q_\beta$'s are the lowering and the $\bar{Q}_\beta$'s are the raising operators. The vacuum $|\Lambda\rangle$ is introduced 
by requiring
\bea
\label{143lowestweightvacua}
&&Q_\beta\left | \Lambda \right \rangle = 0,\qquad\qquad \beta=0,1,2,3.
\eea
From the definition of the $Q_\beta$'s in \eqref{N=4trigcharges} the four differential equations \eqref{143lowestweightvacua} can be recasted into the single differential equation
\begin{equation}
\label{143vacuumsubspace}
(\hat{p} - \frac{i}{2}\hat{y}-\frac{i(1+2\alpha)}{2\hat{y}}\mathcal{F}_4)\left | \Lambda \right \rangle = 0.
\end{equation}
In position-space representation, (\ref{143vacuumsubspace}) splits into two separate equations for  the bosonic ($+$) and respectively  fermionic ($-$) subspaces,
\begin{equation}
\label{143vacuumequation}
\frac{d\phi_{0,\sigma}}{dy} = -\frac{1}{2}(y \pm \frac{1+2\alpha}{y} )\phi_{0,\sigma}.
\end{equation} 
The label $\sigma$ accounts, just as in the parabolic case, for the $\hat{J}$,$\hat{L}$-spin degrees of freedom. \par
Integrating the above equation we get, up to normalization, the vacuum solutions
\bea
\label{143vacuusolutions}
\phi_{0,\sigma} &=& y^{\mp(\frac{1+2\alpha}{2})}\exp(-\frac{y^2}{4}).
\eea
This result is in agreement with \eqref{143trigeigenfuctions} provided that we set
\begin{equation}
\label{alpha-nu-relation-trig}
\nu_b = -(1+\alpha), \qquad \nu_f = \alpha.
\end{equation}
This analysis forces us to conclude that two degenerate lowest energy vacua exist for
$\alpha\neq -\frac{1}{2}$. They are bosonic for $\alpha < -\frac{1}{2}$ and fermionic for $\alpha > -\frac{1}{2}$. This is implied by equation \eqref{143trigeigenfuctions} which tells us that any bosonic (fermionic)  vacuum should be such that $\nu_b +\frac{1}{2}>0$ ($\nu_f +\frac{1}{2}>0$).
\par
At the special $\alpha = -\frac{1}{2}$ value we have that
$D(2,1;-\frac{1}{2})\equiv D(2,1)\approx osp(4|2)$. The Calogero potential terms vanish both in the upper and lower diagonal blocks. 
At $\alpha = -\frac{1}{2}$ we recover four undeformed harmonic oscillator equations. All the states of the theory (including the minimal energy states) are four times degenerated, with two bosonic and two fermionic states of same energy.\par
The energy levels of the system are given by
\bea
\label{143trigenergies-bos-fer}
E_{b,n} =\frac{1}{2}(n -\alpha),&&  E_{f,n} =\frac{1}{2}(n + \alpha+1), \quad n=0,1,2, \ldots .
\eea
$E_{b,n}$ is the whole spectrum of energies recovered from a bosonic vacum ($\alpha < - \frac{1}{2}$). Conversely, $E_{f,n}$ is the whole spectrum when the vacuum is fermionic ($\alpha > \frac{1}{2}$). \par
For a bosonic (fermionic) vacuum, the energy of the two degenerate vacua is respectively given by
\bea
\label{143trigenergies-bos-fer}
E_{b,vac} =-\frac{1}{2} \alpha,\quad  (\alpha\leq-\frac{1}{2})  &;&  E_{f,vac} =\frac{1}{2}( \alpha+1) ,\quad  (\alpha\geq-\frac{1}{2})  .
\eea
The scaling dimension $\alpha$ can be regarded as an external control parameter of the theory, so that the vacuum energy can be interpreted as a Casimir energy. The Casimir energy of the $(1,4,3$) $D(2,1;\alpha)$ (un)deformed oscillator admits a very nice expression in terms of $\alpha$, being simply given by
\bea
\label{143trigenergies-alpha}
E_{vac} &=& \frac{1}{4}(1+|2\alpha+1|).
\eea

This expression should be compared with the much more complicated expression of the vacuum energy in terms of the Calogero coupling constant $g$ and derived from (\ref{g-nu-relation-trig}). This result suggests that the scaling dimension $\alpha$ has a more direct physical interpretation of the Calogero coupling constant $g$. One should also note that, contrary to $g$,  $\alpha$ directly enters the spectrum-generating  superalgebra
$D(2,1;\alpha)$.

\subsection{The $\mathcal{N}=2$ $(2,2,0)$ trigonometric model with $sl(2|1)$ invariance}

As in the parabolic case, we obtain from quantization the non-vanishing (anti)commutators
\bea
\label{com}
[y^{*},p_{y*}]=[y,p_y]=i\hbar,&\quad&\{ \chi,\chi^{\dagger}\} =\frac{\hbar}{\omega C},
\eea
with $\chi=\sqrt{\frac{\hbar}{\omega C}}\left(\begin{array}{cc}
0 & 1\\
0 & 0
\end{array}\right)$ and $\chi^{\dagger}=\sqrt{\frac{\hbar}{\omega C}}\left(\begin{array}{cc}
0 & 0\\
1 & 0
\end{array}\right)$.
We work with $\hbar=1$, $C=\frac{1}{2}$, $\omega=2$. Therefore, the quantum operators of the superalgebra can be written as

\bea
\label{opsl21}
\hat{H} &=& i\frac{e^{-2it}}{2}(\hat{{\cal H}}-yy^{*}\mathbb{I}_2+i(y^{*}p_{y^{*}}+yp_{y}-i)\mathbb{I}_2), \nonumber\\
\hat{D} &=& \frac{i}{2}(2p_{y}p_{y^{*}}+\frac{yy^{*}}{2}+\frac{(2\lambda +1)^{2}}{8yy^{*}})\mathbb{I}_2+i\frac{2\lambda +1}{4}(\chi\chi^{\dagger}-\chi^{\dagger}\chi)(\frac{p_{y^{*}}}{y}-\frac{p_y}{y^{*}})=\frac{i}{2}\hat{{\cal H}}, \nonumber\\
\hat{K} &=& i\frac{e^{2it}}{2}(\hat{{\cal H}}-yy^{*}\mathbb{I}_2-i(y^{*}p_{y^{*}}+yp_{y}-i)\mathbb{I}_2),\nonumber\\
{\hat{Q}^{(1)}}_{\pm} &=& -ie^{\mp it}[\frac{1}{2}((\frac{y}{y^{*}})^{\frac{1+2\lambda}{2}}p_{y} + p_{y}(\frac{y}{y^{*}})^{\frac{1+2\lambda}{2}})\chi - \frac{1}{2}((\frac{y^{*}}{y})^{\frac{1+2\lambda}{2}}p_{y^{*}} + p_{y^{*}}(\frac{y^{*}}{y})^{\frac{1+2\lambda}{2}})\chi^{\dagger} \nonumber\\
& & \mp \frac{i}{2}(yy^{*})^{\frac{1}{2}}((\frac{y}{y^{*}})^{\lambda}\chi - (\frac{y^{*}}{y})^{\lambda}\chi^{\dagger} )], \nonumber\\
{\hat{Q}^{(2)}}_{\pm}  &=& e^{\mp it}[\frac{1}{2}((\frac{y}{y^{*}})^{\frac{1+2\lambda}{2}}p_{y} + p_{y}(\frac{y}{y^{*}})^{\frac{1+2\lambda}{2}})\chi + \frac{1}{2}((\frac{y^{*}}{y})^{\frac{1+2\lambda}{2}}p_{y^{*}} + p_{y^{*}}(\frac{y^{*}}{y})^{\frac{1+2\lambda}{2}})\chi^{\dagger} \nonumber\\
& & \mp \frac{i}{2}(yy^{*})^{\frac{1}{2}}((\frac{y}{y^{*}})^{\lambda}\chi + (\frac{y^{*}}{y})^{\lambda}\chi^{\dagger} )], \nonumber\\
\hat{J} &=& \frac{i}{2}(\frac{p_{y^{*}}}{y}-\frac{p_y}{y^{*}})-\frac{1-2\lambda}{8}(\chi\chi^{\dagger}-\chi^{\dagger}\chi).
\eea

 \par
The fermionic operators $\hat{Q}_{\pm}^{(I)}$, $I=1,2$, entering $sl(2|1)$, can also be expressed as
\begin{equation}
\hat{Q}_{\pm}^{(1)}=ie^{\mp it}\left(\begin{array}{cc}
0 & -A_{\pm}\\
B_{\pm} & 0
\end{array}\right),\quad\hat{Q}_{\pm}^{(2)}=e^{\mp it}\left(\begin{array}{cc}
0 & A_{\pm}\\
B_{\pm} & 0
\end{array}\right),\label{eq:Q2}
\end{equation}
where, using the polar coordinates as in the parabolic case, we have
\begin{eqnarray}\label{4pi}
A_{\pm} & = & -\frac{i}{2}e^{i2\lambda\theta}(\partial_{r}-\frac{i}{r}\partial_{\theta}+\frac{2\lambda+1}{2r}\pm r),\nonumber \\
B_{\pm} & = & -\frac{i}{2}e^{-i2\lambda\theta}(\partial_{r}+\frac{i}{r}\partial_{\theta}+\frac{2\lambda+1}{2r}\pm r).\label{eq:AAdagger2}
\end{eqnarray}
In the trigonometric case the Hamiltonian $\hat{{\cal H}}$ is related to the Cartan generator ${\hat D}$. We have $\hat{{\cal H}}=-2i\hat{D}$, so that
\begin{equation}\label{dtri}
\hat{{\cal H}}=[-\frac{1}{2}(\partial_{r}^{2}+\frac{1}{r}\partial_{r}+\frac{1}{r^{2}}\partial_{\theta}^{2})+i\frac{\left(2\lambda+1\right)}{2r^{2}}\sigma_{3}\partial_{\theta}+\frac{\left(2\lambda+1\right)^{2}}{8r^{2}}+\frac{r^{2}}{2}]{\mathbb I}_2.
\end{equation}
In the r.h.s. $\sigma_3$ is the diagonal Pauli matrix.\par
For later use we also write the operator $\hat{J}$ as a differential operator,
\bea\label{threefortri}
\hat{J}&=&-\frac{i}{2}{\mathbb I}_2\partial_{\theta}-\frac{2\lambda-1}{4}\sigma_{3}.
\eea
One can check that the $sl(2|1)$ superalgebra is recovered from the (anti)commutators of the operators  (\ref{opsl21}) using (\ref{com}).\par
The differential equation for the radial part of the eigenfunctions $\psi=e^{im\theta}R_{\pm}(r)e_{\pm}$ of $\hat{{\cal H}}$, where $e_{+}={\tiny{\left(\begin{array}{c}
1\\
0
\end{array}\right)}}$ and $e_{-}={\tiny{\left(\begin{array}{c}
0\\
1
\end{array}\right)}}$,  is
\bea
[-\frac{1}{2}(\partial_{r}^{2}+\frac{1}{r}\partial_{r})+\frac{1}{2r^{2}}(m\mp\frac{2\lambda+1}{2})^{2}+\frac{r^{2}}{2}-E]R_{\pm}(r)&=&0.
\eea
$E$ is the energy. In \cite{cal} the same equation is found and solved for the problem of three bodies in a line. Furthermore, the issue of selfadjointness of the differential operator acting on $R_{\pm}$ 
was investigated in \cite{bgms}; since $\sqrt{\left(m\pm\frac{2\lambda+1}{2}\right)^{2}}\geq 0$,
the existence of a selfadjoint extension for the Halmiltonian (\ref{dtri}) is ensured.\par

The requirement of single-valuedness for the operators ${\hat Q}_{\pm}^{(I)}$ on the 
${\mathbb R}^2$-plane implies, from the exponents in (\ref{4pi}), that the constraint $4\lambda \pi=2k\pi$, with $k$ integer, must be satisfied. Therefore the scaling dimension $\lambda$ has to be quantized, either $\lambda=\frac{1}{2} +{\mathbb Z}$ or
$\lambda ={\mathbb Z}$. We discuss in detail the half-integer case, with side remarks about the
models with integer values of $\lambda$. \par
One should note that at $\lambda=-\frac{1}{2}$ one obtains (two copies of) the Hamiltonian of the undeformed two-dimensional bosonic oscillator.\par
For half-integer $\lambda$ the ${\hat Q}_{\pm}^{(I)}$ operators act as raising/lowering operators.  Let us take, e.g., 
$\hat{Q}_{\pm}^{(2)}$; it follows, from the commutators $[\hat{{\cal H}},\hat{Q}_{\pm}^{(2)}]=\mp\hat{Q}_{\pm}^{(2)}$,
that an energy eigenstate $\psi$ with eigenvalue $E_n$ is mapped into an eigenstate $\hat{Q}_{\pm}^{(2)}\psi$
with eigenvalue $E_{n}\mp 1$ (provided that $E_n\mp 1\neq 0$):\par
 $\hat{{\cal H}}\psi=E_{n}\psi\rightarrow \hat{{\cal H}}\hat{Q}_{\pm}^{(2)}\psi=\left(E_{n}\mp1\right)\hat{Q}_{\pm}^{(2)}\psi$.\par
Therefore, starting from a lowest weight state satisfying $\hat{Q}_{+}^{(2)}\psi=0$, an infinite tower of higher energy eigenstates are constructed by repeatedly applying ${\hat Q}_-^{(2)}$.
The solutions of the lowest weight equation $\hat{Q}_{+}^{(2)}\psi=0$ are given by the eigenfunctions
\begin{eqnarray}\label{lwstates}
\psi_{m+}\left(r,\theta\right) & = & A_{m}r^{(m-\frac{2\lambda+1}{2})}e^{-r^{2}}e^{im\theta}\left(\begin{array}{c}
1\\
0
\end{array}\right),\nonumber \\
\psi_{m-}(r,\theta) & = & B_{m}r^{-(m+\frac{2\lambda+1}{2})}e^{-r^{2}}e^{im\theta}\left(\begin{array}{c}
0\\
1
\end{array}\right),
\end{eqnarray}
where $A_{m}$, $B_{m}$ are normalization constants given by%
\begin{eqnarray}
A_{m}&=&2^{\frac{\alpha+1}{2}}\frac{1}{\sqrt{\pi\Gamma (\alpha+1)}} ,\quad \alpha=m-\frac{2\lambda+1}{2},\nonumber\\
 B_{m}&=&2^{\frac{\beta+1}{2}}\frac{1}{\sqrt{\pi\Gamma (\beta+1)}},\quad\beta=-(m+\frac{2\lambda+1}{2})
\end{eqnarray}
and $\Gamma$ is the gamma function. \par
In order to have finite lowest weight eigenfunctions at the origin, the integer $m$ is constrained. From the bosonic states the necessary condition is 
\bea \label{ineq1} 
m&\geq&\frac{2\lambda+1}{2},
\eea 
while from the fermionic states the necessary condition is 
\bea\label{ineq2}
m&\leq&-\frac{2\lambda+1}{2}.
\eea
The energy eigenvalue equation  of the bosonic and fermionic lowest weight eigenstates is respectively given by
\begin{eqnarray}
\hat{{\cal H}}\psi_{m+} & = & (1+m-\frac{2\lambda+1}{2})\psi_{m+},\nonumber \\
\hat{{\cal H}}\psi_{m-} & = & (1-(m+\frac{2\lambda+1}{2}))\psi_{m-}.
\end{eqnarray}
Two minimal vacua, one bosonic and the other fermionic, are obtained with vacuum energy $1$. They are recovered from the ``saturated" bosonic and fermionic lowest weight eigenstates with, respectively,
$m=\frac{2\lambda+1}{2}$ and $m=-\frac{2\lambda+1}{2}$.\par
The same set of lowest weight states given by formula (\ref{lwstates}) is obtained from the lowest weight condition associated with the lowering operator $Q_+^{(1)}$  ($Q_+^{(1)}\psi=0$). The repeated application of the raising operator $Q_-^{(1)}$ applied to a lowest state reconstructs, up to a phase, the higher energy states obtained from the raising operator $Q_-^{(2)}$.\par
The theory therefore possesses a degenerate vacuum, one vacuum state being bosonic, the other one fermionic. As discussed in Appendix {\bf A} it is possible to impose a superselection rule, imposed by a projector, which selects half of the states being physical. The superselected theory possesses a unique bosonic vacuum and, for $\lambda=-\frac{1}{2}$, its spectrum coincides with the spectrum of the ordinary two-dimensional (undeformed) oscillator, which can therefore be recovered as the superselected, $\lambda=-\frac{1}{2}$, $sl(2|1)$ acting on $(2,2,0)$, quantum trigonometric model.\par
We conclude this Section with two important remarks. Contrary to the two vacua of the (not superselected) $\lambda=\frac{1}{2}+{\mathbb Z}$ theory, the $\lambda\in {\mathbb Z}$ quantum deformed oscillators  possess four
vacuum states (two bosonic and two fermionic states). The construction of the Hilbert space follows the same lines as the half-integer $\lambda$ case. The main difference lies in the fact that the necessary conditions (\ref{ineq1}) and (\ref{ineq2}) for the integer $m$ cannot be satisfied as equalities when $\lambda\in {\mathbb Z}$. It is beyond the scope of this work to present the detailed analysis of the $\lambda\in{\mathbb Z}$ deformed oscillators, which will be presented elsewhere.\par
The second important remark concerns the fact that, for the superselected $\lambda=\frac{1}{2}+{\mathbb Z}$ theory, the Hilbert space cannot be recovered by repeatedly acting with the $sl(2|1)$ raising operators from the vacuum state. The Hilbert space is decomposed (this point is discussed in Appendix {\bf A})
in a infinite direct sum of the $sl(2|1)$ lowest weight representations. This is in sharp contrast with respect to the one-dimensional harmonic oscillator, whose single irreducible lowest weight representation of the $osp(1|2)$ spectrum-generating superalgebra allows to recover the whole Hilbert space.\par
One can note, however, that it is possible to construct an extra set of fermionic symmetry operators, ${\overline Q}_\pm^{(I)}$, which also act
as raising/lowering operators. The construction goes as follows. At first a discrete symmetry operator $\hat{C}$, playing the role of a charge conjugation operator, is introduced. It is given by
\bea
\hat{C}&=&\left(\begin{array}{cc}
0 & e^{i\left(2\lambda+1\right)\theta}\\
e^{-i\left(2\lambda+1\right)\theta} & 0
\end{array}\right).
\eea
One can verify that $[\hat{{\cal H}},\hat{C}]=0$, where $\hat{\cal H}$ is given in (\ref{dtri}), and that $\hat{C}^{2}=\mathbb{I}_2$. The operator  ${\hat C}$ also commutes with the
${\hat{K}}$ and ${\hat{H}}$ operators in (\ref{opsl21}). It does not commute, however, with ${\hat J}$ and the $sl(2|1)$ fermionic operators.\par
With the help of ${\hat C}$ we can introduce the new symmetry operators 
\bea\label{overlineqs}
\hat{C}\hat{Q}_{\pm}^{(1)}\hat{C}= {\overline{Q}}_{\pm}^{(1)}=ie^{\mp it}\left(\begin{array}{cc}
0 & C_{\pm}\\
-D_{\pm} & 0
\end{array}\right),&&\quad\hat{C}\hat{Q}_{\pm}^{(2)}\hat{C}={\overline{Q}}_{\pm}^{(2)}=e^{\mp it}\left(\begin{array}{cc}
0 & C_{\pm}\\
D_{\pm} & 0
\end{array}\right),
\eea
where
\begin{eqnarray}
C_{\pm} & = & -\frac{i}{2}e^{i2\left(\lambda+1\right)\theta}(\partial_{r}+\frac{i}{r}\partial_{\theta}-\frac{2\lambda+1}{2r}\pm r),\nonumber \\
D_{\pm} & = & -\frac{i}{2}e^{-i2\left(\lambda+1\right)\theta}(\partial_{r}-\frac{i}{r}\partial_{\theta}-\frac{2\lambda+1}{2r}\pm r),\label{eq:BBdagger2}
\end{eqnarray}
and
\bea
&\hat{C}\hat{J}\hat{C}={\overline{J}}=-\frac{i}{2}\partial_{\theta}-\frac{2\lambda+3}{4}\sigma_{3}.&
\eea

\par
~\par

Let us collectively denote as ${\hat g}_i$ ($i=1,2,\ldots, 8$) the $sl(2|1)$ operators entering (\ref{opsl21}).
By construction, the operators ${\bar g}_i= {\hat C} {\hat g}_i {\hat C}^{-1}$, obtained through a similarity transformation, close as well the $sl(2|1)$ superalgebra. It is worth pointing out that this second set of $sl(2|1)$ operators cannot be expressed as a linear combination of the ${\hat g}_i$ set of $sl(2|1)$ operators. In particular the (anti)commutators $[{\hat g}_i, {\bar g}_j\}$ produce new operators in the right hand side. It is not clear
which algebraic structure is induced by the combined set of ${\hat g}_i$ and ${\bar g}_j$ operators (see the comments in the Conclusions). An important feature, discussed in Appendix {\bf A}, is the fact that we need rasing operators from both sets, ${\hat g}_i$ and ${\bar g}_j$, to produce {\em every} excited state of the theory 
by applying raising operators on the ground state(s). An exemplification of this is illustrated, e.g., by the 
Figure $1$ diagram of Appendix {\bf A}.  Both ${\hat{Q}}_{\pm}^{(I)}$ and ${\overline{Q}}_{\pm}^{(I)}$ act
as rasing/lowering operators. The action of the ${\hat{Q}}_{\pm}^{(I)}$ raising operators is illustrated by the
solid edges, while the action of the ${\overline{Q}}_{\pm}^{(I)}$ raising operators is illustrated by the dashed edges.\par
In terms of ${\hat C}$ we can also introduce the new quantum operators
\bea\label{calj}
&{\mathcal{J}}=\hat{J}+{\overline{J}}=-i\partial_{\theta}-\frac{2\lambda+1}{2}\sigma_{3},\quad
N_f=\sigma_{3}=\hat{J}- {\overline{J}},&
\eea
which allows us to define the new quantum numbers (used in Appendix {\bf A}, see Figure $4$):
\bea
&\hat{{\cal H}}\left|n,j,\epsilon\right\rangle =\left(n+1\right)\left|n,j,\epsilon\right\rangle,\quad
{\mathcal{J}}\left|n,j,\epsilon\right\rangle =j\left|n,j,\epsilon\right\rangle,\quad
\sigma_{z}\left|n,j,\epsilon\right\rangle =\epsilon\left|n,j,\epsilon\right\rangle.&
\eea

\section{Conclusions}

In this paper we presented a framework for quantizing the large class of classical  worldline superconformal $\sigma$-models derived from supermultiplets. These systems are defined in \cite{kuto} (for the parabolic case) and
\cite{hoto} (for the trigonometric case).  We applied the quantization prescription to derive explicitly the ${\cal N}=4$
$(1,4,3)$ and the ${\cal N}=2$ $(2,2,0)$  quantum superconformal mechanics (with $D(2,1;\alpha)$ and $sl(2|1)$ dynamical symmetry, respectively). The parameter $\alpha\neq0,-1$ is the scaling dimension of the $(1,4,3)$ supermultiplet, while the scaling dimension of the $(2,2,0)$ supermultiplet is quantized and given by $\lambda=\frac{1}{2}+{\mathbb Z}$ or $\lambda\in {\mathbb Z}$. \par
The results concerning the trigonometric models are particularly relevant. These systems are only ``softly supersymmetric", see the discussion in Appendix {\bf C}. As such they have not received much attention like the parabolic models. The trigonometric models  correspond to superconformal mechanics in the presence of the DFF damping oscillatorial term; stated otherwise, they are oscillators where Calogero potential terms are
possibly present. Their spectrum is discrete and bounded from below.\par
For the $(1,4,3)$ trigonometric models (i.e., the $D(2,1;\alpha)$ oscillators) we derive  the following nice formula for the vacuum energy:
\bea
\label{cas}
E_{vac} &=& \frac{1}{4}(1+|2\alpha+1|).
\eea
If $\alpha$ is interpreted as a physical external parameter, then (\ref{cas}) can be interpreted as a Casimir energy.
\par
A restriction (obtained by imposing a superselection rule derived by a projector, see Appendix {\bf A}),
of the $(2,2,0)$ trigonometric model at the special value $\lambda=-\frac{1}{2}$ allows to recover the spectrum of the ordinary two-dimensional oscillator.\par
The (unrestricted)  ${\cal N}=2$} $(2,2,0)$ trigonometric models for the $\lambda\in\frac{1}{2}+{\mathbb Z}$ and $\lambda\in {\mathbb Z}$ quantized values of the scaling dimension possess an $sl(2|1)$ dynamical symmetry. As a consequence, their spectrum is a direct sum of an infinite tower of $sl(2|1)$ lowest weight representations.\par
The surprising presence of an extra fermionic symmetry (discussed at length in Section {\bf 6} and in Appendix {\bf A} and {\bf C}) produces extra fermionic generators which act as raising and lowering operators. They allow to reach each state belonging to the Hilbert space of the two-dimensional models by repeatedly applying the raising operators to the vacuum state.\par
This result seems to suggest the existence of a broader dynamical symmetry algebra (not necessarily a superalgebra, it could be, see \cite{aktt}, a ${\mathbb Z}_2\times {\mathbb Z}_2$-graded dynamical symmetry algebra) which has to be introduced in order to recover the spectrum of the  ${\cal N}=2$ $(2,2,0)$ (deformed) oscillators from a single, irreducible, lowest weight representation. We are planning to address this remarkable feature in our forthcoming investigations.

{~}
\par
{~}\par{}
%\newpage
\renewcommand{\theequation}{A.\arabic{equation}}
\setcounter{equation}{0}
 {~}\\

{\Large{\bf Appendix A: Diagrams of the spectrum-generating superalgebra for the ${\cal N}=2$,  $(2,2,0)$, $\lambda=\frac{1}{2}+{\mathbb Z}$ trigonometric cases.}}\par
{~}~\par

It is convenient, for the two-dimensional cases based on the ${\cal N}=2$ $(2,2,0)$ trigonometric reps, to encode in diagrams  the action of the raising and lowering operators of the spectrum-generating superalgebra. We explicitly present three such diagrams, Figures $1$, $2$ and $3$, respectively associated with three values of the scaling dimension, $\lambda=\frac{1}{2}$, $\lambda=-\frac{1}{2}$, $\lambda=-\frac{3}{2}$. In a further diagram the general features of the $\lambda=\frac{1}{2}+{\mathbb Z}$
case are presented.
\par
In the diagrams the bosonic (fermionic) states are denoted by white (black) dots. Grey dots denote the presence of both bosonic and fermionic states. The vertical axis represents the energy level, labeled by $n$,
while the horizontal axis represents the angular momentum, labeled by $m$. We denote with $\epsilon$ the eigenvalues of the Fermion Number operator ($\epsilon = +1$ for bosons, $\epsilon =-1$ for fermions).
Solid (dashed) lines represent states connected by ${\widehat Q}_\pm^{(I)}$ (respectively, ${\overline Q}_\pm^{(I)}$) raising and lowering operators with $I=1,2$, see (\ref{eq:Q2}) and (\ref{overlineqs}) (for simplicity we drop here the indices).\par 
The $sl(2|1)$ lowest weight states appear, in the diagrams, as the dots where the solid lines originate (in the upward direction). In Figure $2$ and $4$ the existence of such lowest weight states is not immediately evident, this is however just a side effect of the condensed notation used (a grey dot being associated with two states).\par
The operators 
${\widehat{Q}}_\pm ^{(1)}, {{\widehat{Q}_{\pm}^{(2)}}}$ (and, similarly,
${{{\overline{Q}}_{\pm}^{(1)}}},{{{\overline {Q}}_{\pm}^{(2)}}}$),
applied to a $\left|n,m,\epsilon\right\rangle $ state which does not coincide with a lowest weight state produce, apart a normalization factor,  the same state. We can write, for $I=1,2$, 
\begin{eqnarray}
\widehat{Q}_{\pm}^{(I)}\left|n,m,\epsilon\right\rangle &\propto&\left|n\mp1,m-\epsilon2\lambda,-\epsilon\right\rangle ,\nonumber\\
{\overline{Q}}_{\pm}^{(I)}\left|n,m,\epsilon\right\rangle &\propto&\left|n\mp1,m-\epsilon2\left(\lambda+1\right),-\epsilon\right\rangle .
\end{eqnarray}

From the three diagrams, Figures $1$, $2$ and $3$, we can immediately read several important features. In particular, in all three cases, the $n>0$ higher energy states are produced via repeated applications of  the ${\widehat{Q}}$'s, ${\overline Q}$'s raising operators from the two (one bosonic and one fermionic) $n=0$ fundamental level states. 
 As a corollary, we need both types (${\widehat Q}$'s, ${\overline Q}$'s) of raising operators to recover  the Hilbert space of the associated model. This means, stated otherwise, that the Hilbert space is {\em reducible} with respect to the $sl(2|1)$ superalgebra defined by the ${\widehat Q}_\pm^{(I)}$ operators alone. In terms of a $sl(2|1)$ decomposition, an infinite tower (one state at each given integer value  $n$) of lowest weight states need  to be introduced to recover the Hilbert space of the theory.  Therefore, in order to have an irreducible description, the ${\overline Q}_+^{(I)}$ operators need to enter the picture.\par
One shoud note that the $\lambda=-\frac{1}{2}$ case corresponds to the undeformed (namely, without the extra Calogero potential term)  two-dimensional harmonic oscillator. The Hilbert space defined by Figure $2$ contains a double degeneracy. Two eigenstates (one bosonic, the other one fermionic) are associated with each $n,m$ pair of eigenvalues. The introduction of a suitable projection allows to remove the double degeneracy and recover the Hilbert space of the ordinary two-dimensional harmonic oscillator. The superselection rule is defined in terms of the projection operator $\hat{P}$ ($\hat{P}^{2}={\mathbb I}$), given by
\bea\label{supersel}
\hat{P}&=&N_fe^{i\pi\cal{H}},
\eea
where $N_f$ is the Fermion Parity operator and $\hat{{\cal{H}}}=-2i\hat{D}$  is the Hamiltonian (its eigenvalues are the non-negative integers $n$). The 
\bea {\hat P}|\Psi\rangle &=& |\Psi \rangle
\eea 
superselection rule implies that
the Hilbert space of the superselected theory is given by bosonic states at even energy eigenvalues $(n=2k$, with $k=0,1,2,\ldots$) and fermionic states at odd energy eigenvalues $(n=2k+1)$.\par
The superselection removes, in particular, the degeneracy of the vacuum, the single vacuum state being now bosonic. The spectrum of the ordinary two-dimensional harmonic oscillator is therefore recovered from the {\em superselected} ${\cal N}=2 $ $(2,2,0)$ model at scaling dimension $\lambda=-\frac{1}{2}$.\par
For any half-integer value $\lambda =\frac{1}{2}+{\mathbb Z}$ the Hilbert space of the two-dimensional deformed (due to the presence, besides the quadratic potential,  of a Calogero potential term) harmonic oscillator, can be formally recovered from the $\lambda=-\frac{1}{2}$ Figure $2$ diagram, by replacing the angular momentum $m$ with the $j$ eigenvalues of the ${\cal J}$ operator introduced in (\ref{calj}) (this is also true for the $\lambda=\frac{1}{2},-\frac{3}{2}$ cases explicitly introduced in Figure $1$ and $3$).\par
Let us introduce the basis defined by the quantum numbers
$$
\hat{\cal{H}}\left|n,j,\epsilon\right\rangle =\left(n+1\right)\left|n,j,\epsilon\right\rangle ;\quad
\hat{\mathcal{J}}\left|n,j,\epsilon\right\rangle =j\left|n,j,\epsilon\right\rangle ,
(j\in\mathbb{Z});\quad N_f\left|n,j,\epsilon\right\rangle =\epsilon\left|n,j,\epsilon\right\rangle ,
(\epsilon=\pm1).
$$ 
In this basis the action of $\hat{Q}_{\pm}^{(I)}$, ${\overline{Q}}_{\pm}^{(I)}$ on a state which does not coincide with a lowest weight state,  reads as follows 
\begin{equation}
\hat{Q}_{\pm}^{(I)}\left|n,j,\epsilon\right\rangle \propto\left|n\mp1,j+\epsilon,-\epsilon\right\rangle ,\quad {\overline{Q}}_{\pm}^{(I)}\left|n,j,\epsilon\right\rangle \propto\left|n\mp1,j-\epsilon,-\epsilon\right\rangle .
\end{equation}
The $\lambda=\frac{1}{2}+{\mathbb Z}$ associated diagrams are presented in Figure $4$.\par
This makes clear that the superselection rule induced by (\ref{supersel}) can be imposed on any $\lambda=\frac{1}{2}+{\mathbb Z}$ deformed oscillator, guaranteeing in all these cases the existence of a Hilbert space with a single bosonic vacuum.

\begin{figure}
\begin{centering}
\includegraphics{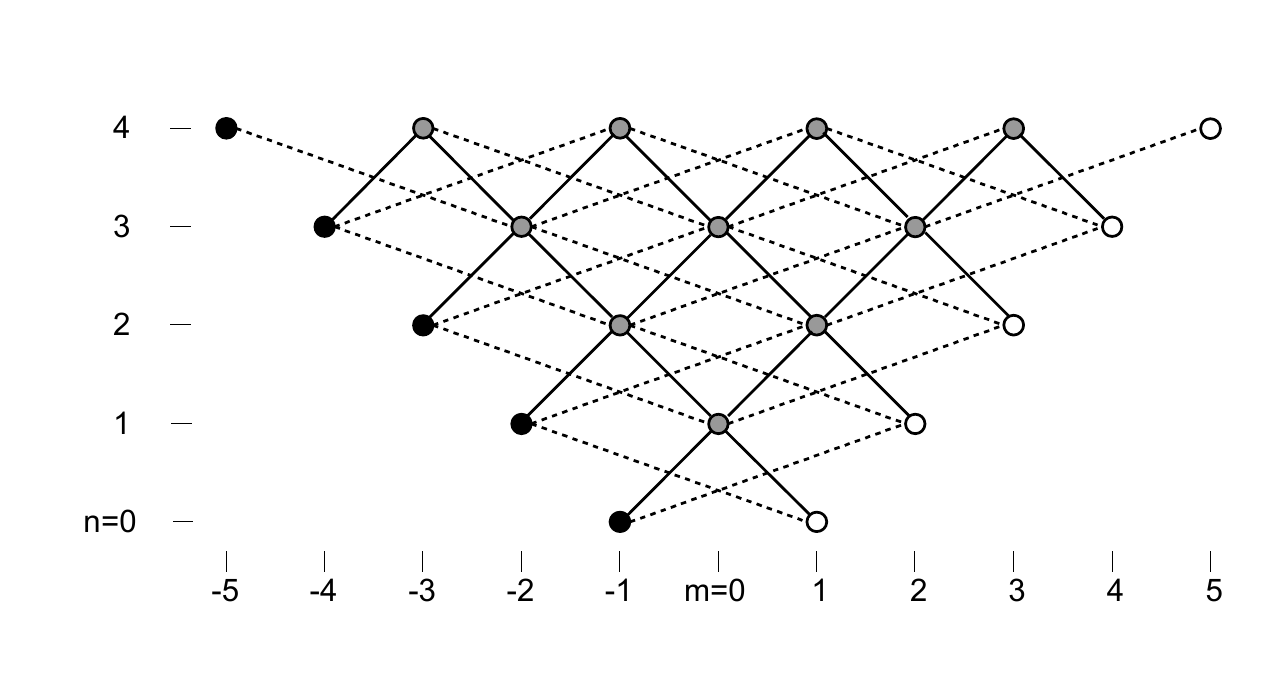}
\protect\caption{$\lambda=\frac{1}{2}$ diagram of ${\widehat Q}$'s, $\overline{Q}$'s raising and lowering operators.}
\end{centering}
\label{fig1}
\end{figure}

\begin{figure}
\begin{centering}
\includegraphics{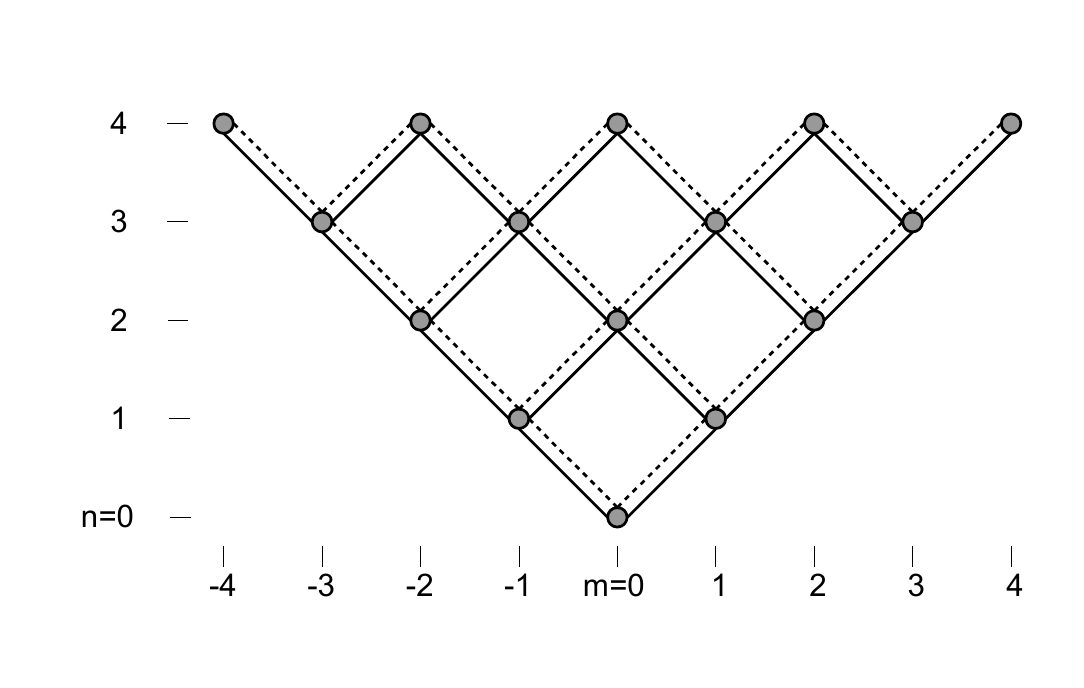}
\protect\caption{$\lambda=-\frac{1}{2}$ diagram of ${\widehat Q}$'s, $\overline{Q}$'s raising and lowering operators..}
\end{centering}
\label{fig2}
\end{figure}

\begin{figure}
\begin{centering}
\includegraphics{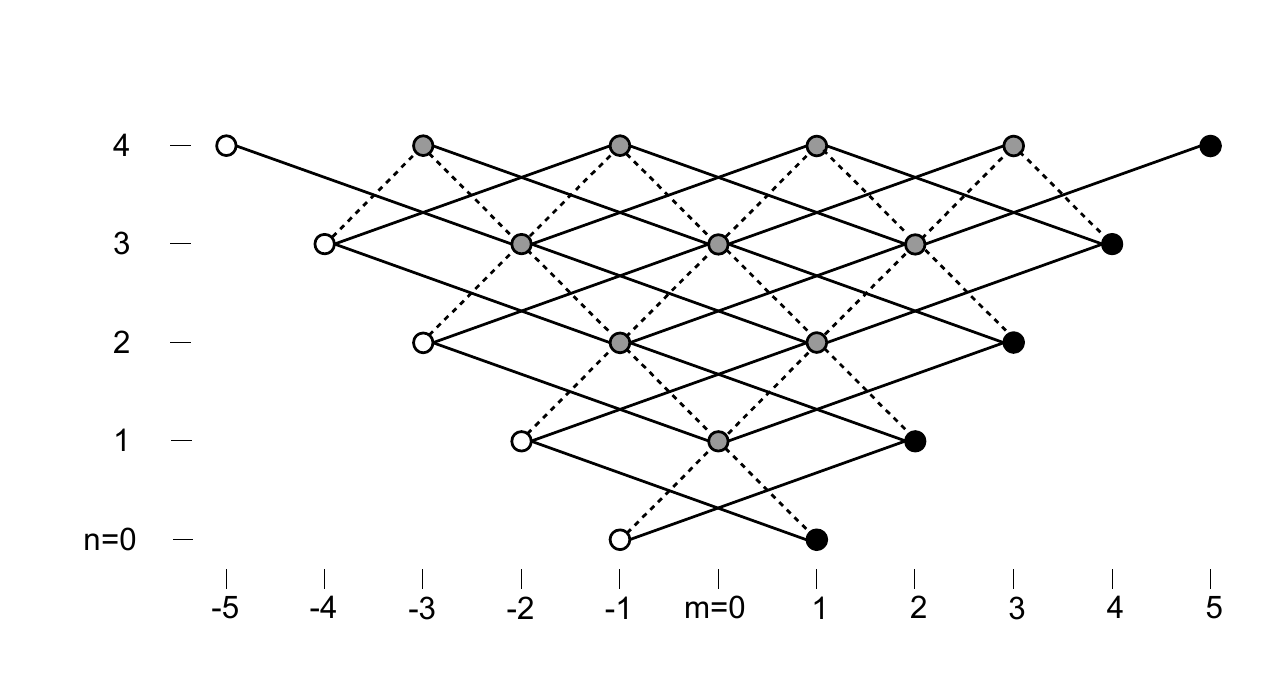}
\protect\caption{$\lambda=-\frac{3}{2}$ diagram of ${\widehat Q}$'s, $\overline{Q}$'s raising and lowering operators.}
\end{centering}
\label{fig3}
\end{figure}

\begin{figure}
\begin{centering}
\includegraphics{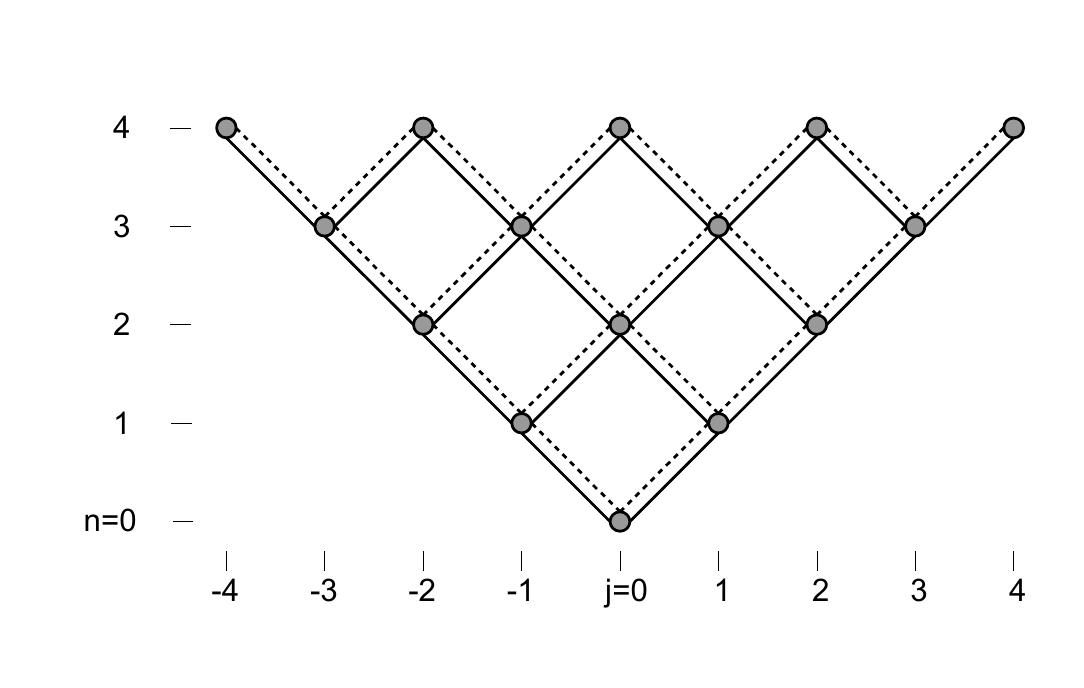}
\protect\caption{the $\lambda=\frac{1}{2}+{\mathbb Z}$ general diagram.}
\end{centering}
\label{fig4}
\end{figure}

{~}
\par
{~}\par{}
\newpage
\renewcommand{\theequation}{B.\arabic{equation}}
\setcounter{equation}{0}
 {~}\\

{\Large{\bf Appendix B: The classical $(2,2,0)$ $sl(2|1)$-invariant models.}}\par
{~}~\par

We present, for completeness, the construction of the $sl(2|1)$-invariant classical actions obtained from, respectively,
the parabolic and the trigonometric $D$-module reps acting on the $(2,2,0)$ supermultiplet.\par
The parabolic D-module rep is given by the transformations
\begin{eqnarray}
&&L_{n}x_{i}=t^{n}\left(t\dot{x}_{i}+\left(n+1\right)\lambda x_{i}\right), \quad \quad L_{n}\psi_{i}=t^{n}(t\dot{\psi}_{i}+\left(n+1\right)(\frac{2\lambda+1}{2})\psi_{i}),\; n=0,\pm1;\nonumber \\
&&Jx_{i}=-\lambda\epsilon_{ij}x_{j},  ~~\qquad\quad\quad\quad\quad\quad\quad J\psi_{i}=-\frac{2\lambda-1}{2}\epsilon_{ij}\psi_{j},\nonumber \\
&&Q_{\pm}^{1}x_{i}=t^{\frac{1\pm1}{2}}\epsilon_{ij}\psi_{j},~~ \qquad\quad\quad\quad\quad Q_{\pm}^{1}\psi_{i}=-it^{\frac{1\pm1}{2}}\epsilon_{ij}\left(t\dot{x}_{j}+\left(1\pm1\right)\lambda x_{j}\right),\nonumber \\
&&Q_{\pm}^{2}x_{i}=t^{\frac{1\pm1}{2}}\psi_{i}, \quad\qquad\quad \quad\quad\quad\quad  Q_{\pm}^{2}\psi_{i}=it^{\frac{1\pm1}{2}}\left(t\dot{x}_{i}+\left(1\pm1\right)\lambda x_{i}\right),\label{eq:transp220}
\end{eqnarray}
\\
where the $x_{i}$'s ($i=1,2)$ are the propagating bosons and the $\psi_{i}$'s the
fermionic fields.\par
 The above transformations close the $sl\left(2|1\right)$ superalgebra.\par
The $sl(2|1)$-invariant action is obtained from the Lagrangian $\mathcal{L}=Q^{2}_+Q^{1}_+(\frac{1}{2}F\epsilon_{ij}\psi_{i}\psi_{j})$, with the operators $Q^{2}_+,Q^{1}_+$ acting on the prepotential $F=C(x_{i}x_{i})^{-\frac{2\lambda+1}{2\lambda}}$ ($C$ is a normalization constant). Explicitly, the invariant action of the classical $(2,2,0)$ parabolic model is
\bea
&{\cal S} =\int dt \mathcal{L}=\int dt(F(\dot{x}_{i}\dot{x}_{i}-i\dot{\psi}_{i}\psi_{i})-iF_{i}\dot{x}_{j}\psi_{i}\psi_{j}).&
\eea
The trigonometric $D$-module rep is given by the transformations
\bea
&&L_{n}x_{i}=\frac{e^{-in\omega t}}{-i\omega}\left(\dot{x}_{i}-in\lambda\omega x_{i}\right), \quad\quad\quad L_{n}\psi_{i}=\frac{e^{-in\omega t}}{-i\omega}(\dot{\psi}_{i}-in(\frac{2\lambda+1}{2})\omega\psi_{i}),\; n=0,\pm1;\nonumber \\
&&Jx_{i}=-\lambda\epsilon_{ij}x_{j},\qquad \qquad \qquad\qquad \qquad  J\psi_{i}=-\frac{2\lambda-1}{2}\epsilon_{ij}\psi_{j},\nonumber \\
&&Q_{\pm}^{1}x_{i}=e^{\mp i\frac{\omega}{2}t}\epsilon_{ij}\psi_{j},\quad~\qquad\qquad \qquad Q_{\pm}^{1}\psi_{i}=\frac{e^{\mp i\frac{\omega}{2}t}}{i\omega}\epsilon_{ij}\left(\dot{x}_{j}\mp i\lambda\omega x_{j}\right),\nonumber \\
&&Q_{\pm}^{2}x_{i}=e^{\mp i\frac{\omega}{2}t}\psi_{i},  ~\qquad\qquad\quad\qquad\quad Q_{\pm}^{2}\psi_{i}=\frac{e^{\mp i\frac{\omega}{2}t}}{-i\omega}\left(\dot{x}_{i}\mp i\lambda\omega x_{i}\right).\label{trig220}
\eea
Without loss of generality we can set $\omega=1$. The classical action, $sl(2|1)$-invariant under the (\ref{trig220}) trigonometric transformations, is therefore given by
\bea
&{\cal S} =\int dt \mathcal{L}=\int dt(F(\dot{x}_{i}\dot{x}_{i}-i\dot{\psi}_{i}\psi_{i})-iF_{i}\dot{x}_{j}\psi_{i}\psi_{j}+ C\lambda^{2}\left(x_{i}x_{i}\right)^{-\frac{1}{2\lambda}}).&
\eea

{~}
%\par
{~}\par{}
%\newpage
\renewcommand{\theequation}{C.\arabic{equation}}
\setcounter{equation}{0}
 {~}\\

{\Large{\bf Appendix C: On the ``soft" supersymmetry of the oscillators.}}\par
{~}~\par

We make here some comments on the role of superalgebras applied to oscillators (either the ordinary quantum oscillators or the oscillators which are ``deformed" by the presence of a Calogero potential term).\par
The starting point is the famous work of Wigner \cite{wig}. In modern terms, after the concept of superalgebra was introduced in mathematics, Wigner's results can be reinterpreted  (see \cite{cct}) according to the following lines.  For the ordinary quantum oscillator, with creation/annihilation operators $a$, $a^\dagger$ (satisfying $[a,a^\dagger]=1$) and symmetrized Hamiltonian ${\cal H}=\{a,a^\dagger\}$,  we can
assign odd-grading to the operators $a,a^\dagger$, so that they belong to a set of $5$ operators, $a, a^\dagger, a^2, (a^\dagger)^2, 
{\cal H}=\{a,a^\dagger\}$, closing the $osp(1|2)$ superalgebra under (anti)commutations. The last three (bosonic) operators close the $sl(2)$ subalgebra. Under this construction we have an alternative point of view for describing the computation of the the spectrum of the ordinary (one-dimensional) harmonic oscillator: 
we can state that, instead of deriving it from the Fock vacuum $|0\rangle$, annihilated by $a$ ($a|0\rangle=0$),
the spectrum is obtained from a lowest weight representation of $osp(1|2)$, the Hamiltonian being the Cartan element. By adopting this viewpoint the superalgebra $osp(1|2)$ becomes a spectrum-generating superalgebra for the ordinary quantum oscillator, with its Hilbert space being recovered from a single, irreducible, $osp(1|2)$ lowest weight representation.\par
One should note that the bosonic $sl(2)$ subalgebra also acts as a spectrum-generating algebra for the harmonic oscillator. The Hilbert space of the harmonic oscillator is, however, reducible under the $sl(2)$ decomposition. It is given by the direct sum of two irreducible $sl(2)$ lowest-weight representations. The first lowest state is the vacuum of the theory (proportional to the gaussian $e^{-x^2}$ under proper conventions and normalization). The other lowest state is the first excited state, with eigenfunction proportional to $xe^{-x^2}$ and having odd-parity with respect to the $x\mapsto -x$ transformation.
The two $sl(2)$ lowest weight reps correspond to, respectively, the even-parity and the odd-parity energy eigenstates. The role of the fermionic operators in $osp(1|2)$ consists in connecting energy eigenstates of even and odd parity.\par
After the introduction and the subsequent classification of simple Lie superalgebras \cite{{kac},{nah}}, the Wigner's 
approach was advocated in \cite{ganpal}, with special emphasis on parastatistics, prompting a series of investigations on lowest weight representations of simple Lie superalgebras (for a recent review see, e.g., \cite{vderj}).\par
On a separate development the DFF ``trick" of introducing oscillator damping potentials in conformal mechanics relates oscillators (with/without the Calogero potential term) to conformal algebras.\par
It was recognized in \cite{pap} that, due to the DFF ``trick", the introduction of new potentials for conformal mechanics became possible. The two aspects, superalgebra versus conformal algebra, were reconciled in \cite{hoto}. The notion of parabolic versus trigonometric/hyperbolic $D$-module reps of superconformal algebras was pointed out, with the latter class describing the (deformed or undeformed) oscillators and bounded from below potentials in the trigonometric case.\par
The main property shared by the two big classes of superconformal theories, parabolic versus trigonometric, is that at the classical level their respective actions are superconformally invariant. Concerning their differences:
\\
{\em i}) the parabolic models are, both classically and quantum, superconformal and supersymmetric. The supersymmetry implies the existence of a symmetry operator ${\cal Q}$ which is the ``square root" of the Hamiltonian ${\cal H}$, namely ${\cal Q}^2={\cal H}$;\\
{\em ii}) the trigonometric models, on the other hand, despite being superconformally invariant, are not supersymmetric. In this case symmetry operators ${\cal Q}$, ${\cal Z}$ exist such that ${\cal Q}^2={\cal Z}$. The key point is that the operator ${\cal Z}$ does not coincide with the Hamiltonian: ${\cal Z}\neq {\cal H}$.\par
One can easily say that the trigonometric models are ``intermediate" between the supersymmetric and the non-supersymmetric theories. This ``intermediate notion of supersymmetry", namely ${\cal Q}^2={\cal Z}\neq {\cal H}$, has no special name in the literature. In \cite{hoto} the notion of ``weak supersymmetry" was employed, borrowing the term from a construction described in \cite{smi} which shares a similar feature.
The use of the term ``weak supersymmetry", however, could be misleading since the models in \cite{smi} are not based on superconformal algebras. In that paper a  ``weak supersymmetric oscillator" is discussed that has no relation with the oscillators derived from the trigonometric $D$-module reps of superconformal algebras. \par
For this reason it seems more appropriate to denote this important class of trigonometric models (which include, as shown in this paper,  the ordinary one-dimensional and two-dimensional harmonic oscillators)
as ``softly supersymmetric". As far as we know the term ``soft supersymmetry" has not been employed in a different context, making this term both suitable and available to describe the special properties of the trigonometric superconformal mechanics.\par
The softly supersymmetric trigonometric models are characterized by\\
{\em i}) classical superconformal invariance of the action;\\
{\em ii}) spontaneous breaking of the superconformal invariance. Indeed, in the simplest application, the Fock vacuum $|0\rangle$ of the harmonic oscillator is annihilated by $a$ and not by the hermitian operator
$a+a^\dagger$: $(a+a^\dagger)|0\rangle\neq0$;\\
{\em iii}) in the quantum case  the role of the superconformal algebra is that of a spectrum-generating superalgebra. \par
Concerning the last point, we indeed proved, see Appendix {\bf A}, that the spectrum of the ordinary two-dimensional oscillator is decomposed into an infinite tower of $sl(2|1)$ irreducible lowest weight representations. The puzzling presence of the extra fermionic generators (\ref{overlineqs}) which connect eigenstates belonging to different lowest weight reps reminds the role, just discussed above, played by the $osp(1|2)$ fermionic generators in connecting the two $sl(2)$ lowest weight reps of the one-dimensional oscillator.

%\newpage

~
\\ {~}~
\par {\Large{\bf Acknowledgments}}
{}~\par{}~\par

N. L. H. acknowledges FAPERJ for support. I. E. C. acknowledges CNPq for support.  F.T. received support from CNPq (PQ Grant No. 306333/2013-9).

\end{document}